\documentclass[onecolumn]{aa}
\usepackage{graphicx}
\usepackage{natbib}
\usepackage{amssymb}
\usepackage{amstext}
\usepackage{latexsym} 
\usepackage[sumlimits]{amsmath}
\newcommand{\Ha}{H$\alpha$}			
\newcommand{\NII}{[N{\sc ii}]}			
\newcommand{\HII}{H{\sc ii}}			

\begin{document}
   \title{The H$\alpha$ Galaxy Survey  
     \thanks{
      Based on observations made with the Jacobus
      Kapteyn Telescope which was operated 
      on the island of La Palma by the Isaac Newton Group in the Spanish 
      Observatorio del Roque de los Muchachos of the Instituto de Astrof\'\i sica 
      de Canarias. }
        }
    \subtitle{III. Constraints on supernova progenitors from spatial
     correlations with \Ha\ emission}
   \author{P. A. James, 
          \and
          J. P. Anderson
           }
 
          \offprints{P. A. James} 

          \institute{Astrophysics Research
	  Institute, Liverpool John Moores University, \\
          Twelve Quays
	  House, Egerton Wharf, Birkenhead CH41 1LD, UK \\
	  \email{paj@astro.livjm.ac.uk} 
          }
          
  \date{Received ; accepted }
 
  \abstract
{}
{We attempt to constrain progenitors of the different types of supernovae
from their spatial distributions
relative to star formation regions in their host galaxies, as traced
by \Ha\ $+$ \NII\ line emission.}
{We analyse 63 supernovae which have occurred within galaxies from our 
\Ha\ survey of the local Universe.
Three statistical tests are used,
based on pixel statistics, \Ha\ radial growth curves, and total galaxy
emission-line fluxes.}
{Many more type II supernovae come from regions of low or zero
emission line flux than would be expected if the latter accurately
traces high-mass star formation. We interpret this excess as a 40\%
`Runaway' fraction in the progenitor stars.  Supernovae of types Ib
and Ic do appear to trace star formation activity, with a much higher
fraction coming from the centres of bright star formation regions than
is the case for the type II supernovae.  Type Ia supernovae overall show 
a weak correlation with locations of current star
formation, but there is evidence that a significant minority, up to about 
40\%, may be linked to the young stellar population.  The radial 
distribution of all core-collapse supernovae
(types Ib, Ic and II) closely follows that of the line emission and
hence star formation in the their host galaxies, apart from a central
deficiency which is less marked for supernovae of types Ib and Ic 
than for those of type II.  Core-collapse supernova rates overall are 
consistent with being proportional to galaxy total luminosities and star formation
rates; however, within this total the type Ib and Ic supernovae show a 
moderate bias towards more luminous host galaxies, and type II supernovae a slight 
bias towards lower-luminosity hosts.}
{}


\maketitle

%
\section{Introduction}
\label{sec:intro}

Recent years have seen an upsurge of interest in supernovae, with the
high-profile application of Type Ia supernovae (SNIa henceforth) as
cosmological standard candles resulting in the establishment of
dark-energy dominated models as the currently preferred cosmology
\citep{riess98}. However, there remain substantial gaps in our
understanding of SNe of all types, which motivate detailed study of
nearby events. One question concerns the nature of the progenitors of
core-collapse (CC) SNe, and in particular how those of SNII differ
from the rarer SNIb and Ic types; do the latter simply have higher
mass progenitors, or are other factors such as chemical abundance
important?  There are also questions regarding the nature of SNIa
progenitors, specifically whether they are single-degenerate (SD)
accreting white dwarfs, or double-degenerate (DD) coalescing white
dwarf pairs.  Interest in this question has been sharpened by several
recent papers.  \citet{stro04} compare the redshift distribution of
SNIa with estimates of the star formation history of the Universe, and
conclude that there must be a delay of 2--4~Gyr between the formation
of a generation of stars and SNIa events associated with that
population.  However, \cite{greg05} uses population
synthesis modelling techniques to predict a rapid peak of SNIa in the
first 500~Myr after a burst of star formation, with the size of the
peak depending on whether single- or double-degenerate SNIa
progenitors are assumed.  \cite{mann06} 
find evidence for two populations of SNIa progenitors, with about 50\% 
being `prompt' SNe with high-mass ($\sim$5.5~M$_{\odot}$) progenitors and 
a delay of only 10$^8$ years after star formation, and the remainder a `tardy'
population with lower-mass progenitors and typical time delays of several 
Gyr after star formation. 

The direct detection of supernova progenitors in archival images has
had some success, with good candidates for the progenitor star being
identified in at least 7 cases \citep[and references therein]{li05}.
However, this can only be done for core-collapse SNe in very nearby
galaxies.  An alternative method for tackling these questions, which
we adopt here, is to look at the environment of SNe, to put
statistical constraints on the nature of progenitors on the basis of
their association with young or old stellar populations.  For example,
\citet{John63} and \citet{maza76}, following an even earlier
suggestion by \citet{reav53}, demonstrated that SNII tend to be
clustered towards spiral arms in galaxy discs.

The utility of \Ha\ emission-line imaging in this context can be made
clear by the
comment by \citet{kenn98}, in a review of \Ha\ techniques, that
`only stars with masses $>10~M_{\odot}$ and lifetimes of $<$20~Myr
contribute significantly to the integrated ionizing flux'. Thus, if
our understanding of this line emission is secure, \Ha\ should trace
very directly the parent population of the core-collapse SNe.
Some such studies have already been undertaken, e.g. by Kennicutt
himself \citep{kenn84}, who found that SNII rates correlate linearly with total
galaxy \Ha\ luminosities, whereas any correlation between SNI rates
and \Ha\ luminosity was less clear.  Only 11 SNII were included in
this study, but it was still possible to put useful constraints on the
lower limit of SNII progenitor masses, at 8$\pm$1~M$_{\odot}$.
\citet{vand92} and \citet{vand96} looked  specifically at the
identification of CC SNe with specific star formation (SF) regions,
within the limitations of existing SN astrometry.  They found most CC
supernovae to lie close to SF regions, with a few exceptions which lay
as much as 40$^{\prime \prime}$ from their nearest region.
\citet{bart94} looked at the distribution of SNe of all types relative
to both their nearest spiral arm using broad-band images, and their
nearest HII region using \Ha\ imaging taken mainly from the
literature, and found that SN of types II and Ib strongly correlated
with both of the features.

The degree of correlation of SNIa with this young stellar population
is much less certain but there may be some useful constraints from an
analysis of the environments of these SNe.  If \citet{stro04} are
correct, one would certainly expect absolutely no correlation between
SNIa locations and sites of active star formation, whereas if
\cite{greg05} and Mannucci (2005) are correct, some fraction of the
SNIa may still be associated with their star formation region.
 
In this paper we use our recent survey of \Ha$+$\NII\ imaging of nearby
galaxies to investigate these questions, and introduce some new
statistical methods to quantify the degree of correlation between
locations of SNe of different types and star forming regions within
these galaxies.


\section{The \Ha\ Galaxy Survey}
\label{sec:hags}

The \Ha\ Galaxy Survey (\Ha GS) is a study of the star formation
properties of a representative sample of galaxies in the local
Universe using the narrow-band imaging technique.  We have imaged 334
nearby galaxies in both the \Ha$+$\NII\ lines and the $R$-band
continuum.  The sample consists of 327 galaxies with Hubble types from
S0/a to Im with recession velocities between 0 and 3000~km~s$^{-1}$,
selected from the Uppsala General Catalogue of Galaxies
\citep{nilson}, hereafter UGC, with the remaining 7 galaxies being
other star-forming galaxies which serendipitously lay in the survey
fields.  All galaxies were observed with the, now unfortunately
decommissioned, 1.0 metre Jacobus Kapteyn Telescope (JKT), part of the
Isaac Newton Group of Telescopes (ING) situated on La Palma in the
Canary Islands.  The selection and the observations of the sample are
discussed in \citet{paper1}, hereafter Paper I. Key elements of this
project are to extend the study of \Ha$+$\NII\ emission properties to fainter
galaxies than has been done in previous large surveys, and to use
well-defined selection criteria reflecting the full mix of
star-forming galaxies present in the UGC.

For the present study, we searched the International Astronomical
Union (IAU) database of supernovae (which is available at
http://cfa-www.harvard.edu/iau/lists/Supernovae.html ) to find all
known supernovae, throughout history, which have occurred within the
327 UGC \Ha GS galaxies.  As of 2005 February 10, there was a total of 63 such
supernovae in 50 galaxies (several hosted more than one supernova). Of
these 63, 12 are classified as type Ia, 8 as types Ib, Ic or Ib/c, 30
as type II and 13 are unclassified.  All types are
taken from the IAU database, along with the positions which we use in
the present paper to locate the SNe relative to the galaxy \Ha$+$\NII\
distributions.

\begin{table*}
\begin{center}
\begin{small}
\begin{tabular}{lrrccrccc}
\hline
\hline
SN  & NGC & UGC & SN Type & SFR & log L$_R$ & NCRPVF(Err) & SFR frac & L$_R$ frac \cr
\hline
2005ad  &  941 & 1954  &  II   & 0.5464 & 9.36  & 0.000 (0.000) &  0.864 & 0.831  \\
2005W   &  691 & 1305  &  Ia   & 0.8810 & 10.12 & 0.000 (0.000) &  0.562 & 0.669  \\  
2005V   & 2146 & 3429  &  Ib/c & 2.6773 & 10.02 & 0.861 (0.005) &  0.091 & 0.033  \\   
2004dg  & 5806 & 9645  &  II   & 1.6406 & 10.06 & 0.554 (0.041) &  0.395 & 0.535  \\  
2004A   & 6207 & 10521 &  II   & 1.6847 & 9.58  & 0.000 (0.000) &  0.660 & 0.729  \\  
2003ie  & 4051 & 7030  &  II   & 1.5093 & 9.75  & 0.373 (0.055) &  0.885 & 0.838  \\  
2003hr  & 2251 & 4362  &  II   & 0.6446 & 9.98  & 0.000 (0.074) &  0.942 & 0.952  \\  
2003Z   & 2742 & 4779  &  II   & 1.5585 & 9.92  & 0.000 (0.035) &  0.736 & 0.675  \\  
2002gd  & 7537 & 12442 &  II   & 1.2383 & 9.48  & 0.167 (0.046) &  0.685 & 0.759  \\  
2002ce  & 2604 & 4469  &  II   & 2.1677 & 9.72  & 0.108 (0.112) &  0.560 & 0.381  \\  
2002bu  & 4242 & 7323  &  --   & 0.1088 & 8.87  & 0.000 (0.000) &  0.930 & 0.896  \\  
2002A   &  --  & 3804  &  II   & 2.2694 & 10.04 & 0.401 (0.170) &  0.253 & 0.419  \\  
2001bg  & 2608 & 4484  &  Ia   & 1.3340 & 9.96  & 0.000 (0.020) &  0.902 & 0.760  \\  
2001X   & 5921 & 9824  &  II   & 3.3419 & 10.22 & 0.975 (0.002) &  0.369 & 0.579  \\  
2000ew  & 3810 & 6644  &  Ic   & 4.2056 & 10.02 & 0.907 (0.004) &  0.147 & 0.261  \\  
2000db  & 3949 & 6869  &  II   & 2.4053 & 9.73  & 0.723 (0.006) &  0.253 & 0.364  \\  
2000E   & 6951 & 11604 &  Ia   & 4.2698 & 10.11 & 0.204 (0.135) &  0.279 & 0.368  \\  
1999el  & 6951 & 11604 &  II   & 4.2698 & 10.11 & 0.709 (0.021) &  0.259 & 0.320  \\  
1999dn  & 7714 & 12699 &  Ib/c & 7.8490 & 9.94  & 0.038 (0.004) &  0.843 & 0.457  \\    
1999br  & 4900 & 8116  &  II   & 2.3140 & 9.80  & 0.099 (0.051) &  0.932 & 0.786  \\  
1998dh  & 7541 & 12447 &  Ia   & 4.4121 & 10.13 & 0.044 (0.031) &  0.449 & 0.465  \\  
1997dt  & 7448 & 12294 &  Ia   & 4.1612 & 10.00 & 0.524 (0.010) &  0.281 & 0.355  \\  
1997dq  & 3810 & 6644  &  Ib   & 4.2056 & 10.02 & 0.296 (0.039) &  0.734 & 0.774  \\  
1997dn  & 3451 & 6023  &  II   & 1.5203 & 9.54  & 0.073 (0.030) &  0.946 & 0.872  \\  
1997db  &  --  & 11861 &  II   & 4.2181 & 9.36  & 0.029 (0.091) &  0.396 & 0.633  \\  
1996ai  & 5005 & 8256  &  Ia   & 3.1988 & 10.49 & 0.615 (0.026) &  0.292 & 0.325  \\  
1996ae  & 5775 & 9579  &  II   & 5.0033 & 10.35 & 0.747 (0.061) &  0.671 & 0.757  \\  
1995ag  &  --  & 11861 &  II   & 4.2181 & 9.36  & 0.660 (0.195) &  0.170 & 0.343  \\  
1995V   & 1087 & 2245  &  II   & 3.2074 & 9.88  & 0.424 (0.031) &  0.497 & 0.368  \\  
1993X   & 2276 & 3740  &  II   & 13.482 & 10.33 & 0.077 (0.037) &  0.619 & 0.899  \\  
1991N   & 3310 & 5786  &  Ib/c & 11.231 & 10.08 & 0.759 (0.002) &  0.277 & 0.268  \\    
1991G   & 4088 & 7081  &  II   & 3.0566 & 9.91  & 0.066 (0.061) &  0.453 & 0.466  \\  
1990U   & 7479 & 12343 &  Ic   & 6.6349 & 10.25 & 0.712 (0.017) &  0.488 & 0.603  \\  
1987K   & 4651 & 7901  &  II   & 2.4632 & 10.09 & 0.746 (0.013) &  0.303 & 0.409  \\  
1984R   & 3675 & 6439  &  --   & 1.0789 & 9.86  & 0.486 (0.055) &  0.366 & 0.392  \\  
1984F   &  --  & 4260  &  II   & 1.3670 & 9.23  & 0.112 (0.099) &  0.705 & 0.736  \\  
1983I   & 4051 & 7030  &  Ib   & 1.5093 & 9.75  & 0.350 (0.095) &  0.473 & 0.498  \\  
1982B   & 2268 & 3653  &  Ia   & 4.1224 & 10.23 & 0.000 (0.022) &  0.572 & 0.507  \\  
1980L   & 7448 & 12294 &  --   & 4.1612 & 10.00 & 0.000 (0.001) &  0.781 & 0.824  \\  
1979B   & 3919 & 6813  &  Ia   & 0.3105 & 9.18  & 0.000 (0.000) &  0.670 & 0.784  \\  
1976G   &  488 & 907   &  --   & 0.7085 & 10.53 & 0.000 (0.005) &  0.922 & 0.926  \\  
1976E   & 7177 & 11872 &  --   & 0.5739 & 9.69  & 0.535 (0.066) &  0.361 & 0.502  \\  
1975E   & 4102 & 7096  &  --   & 1.7388 & 9.90  & 0.475 (0.010) &  0.648 & 0.573  \\  
1974G   & 4414 & 7539  &  Ia   & 0.9977 & 9.70  & 0.000 (0.015) &  0.816 & 0.772  \\  
1974C   & 3310 & 5786  &  --   & 11.231 & 10.08 & 0.433 (0.002) &  0.651 & 0.515  \\  
1971T   & 1090 & 2247  &  --   & 1.2730 & 9.97  & 0.000 (0.054) &  0.385 & 0.469  \\  
1971S   &  493 & 914   &  II   & 1.2032 & 9.69  & 0.174 (0.076) &  0.570 & 0.605  \\  
1969L   & 1058 & 2193  &  II   & 0.3330 & 8.97  & 0.000 (0.000) &  1.000 & 1.000  \\  
1968W   & 2276 & 3740  &  --   & 13.482 & 10.33 & 0.388 (0.033) &  0.069 & 0.076  \\  
1968V   & 2276 & 3740  &  II   & 13.482 & 10.33 & 0.433 (0.032) &  0.790 & 0.699  \\  
1964H   & 7292 & 12048 &  II   & 0.7043 & 9.03  & 0.059 (0.064) &  0.719 & 0.551  \\  
1963J   & 3913 & 6813  &  Ia   & 0.3105 & 9.18  & 0.000 (0.077) &  0.126 & 0.307  \\  
1963I   & 4178 & 7215  &  Ia   & 0.1181 & 8.47  & 0.317 (0.117) &  0.068 & 0.111  \\  
1962Q   & 2276 & 3740  &  --   & 13.482 & 10.33 & 0.507 (0.025) &  0.472 & 0.455  \\  
1962L   & 1073 & 2210  &  Ic   & 1.0953 & 9.38  & 0.000 (0.000) &  0.518 & 0.754  \\  
1962K   & 1090 & 2247  &  --   & 1.2730 & 9.97  & 0.236 (0.157) &  0.494 & 0.571  \\  
1961V   & 1058 & 2193  &  II   & 0.3330 & 8.97  & 0.363 (0.108) &  0.931 & 0.968  \\  
1960L   & 7177 & 11872 &  --   & 0.5739 & 9.69  & 0.477 (0.125) &  0.933 & 0.949  \\  
1954C   & 5879 & 9753  &  II   & 0.6413 & 9.43  & 0.163 (0.083) &  0.511 & 0.615  \\  
1941C   & 4136 & 7134  &  II   & 0.1423 & 8.88  & 0.000 (0.000) &  0.882 & 0.880  \\  
1937C   &  --  & 8188  &  Ia   & 0.0242 & 7.84  & 0.555 (0.245) &  0.340 & 0.272  \\  
1920A   & 2608 & 4484  &  II   & 1.3340 & 9.96  & 0.000 (0.000) &  0.396 & 0.424  \\  
1895A   & 4424 & 7561  &  --   & 0.0410 & 8.33  & 0.000 (0.000) &  1.000 & 0.943  \\     

\hline
\end{tabular}
\caption[]{Data for the 63 SNe occurring in galaxies from the \Ha GS
survey, and for their host galaxies.}
\label{tbl:1}
\end{small}
\end{center}
\end{table*}


\section{Locations of specific SNe relative to star formation regions}
\label{sec:images}

\subsection{Data reduction and astrometric methods}

\begin{figure*}
\centering
\includegraphics[angle=0,width=4.5cm]{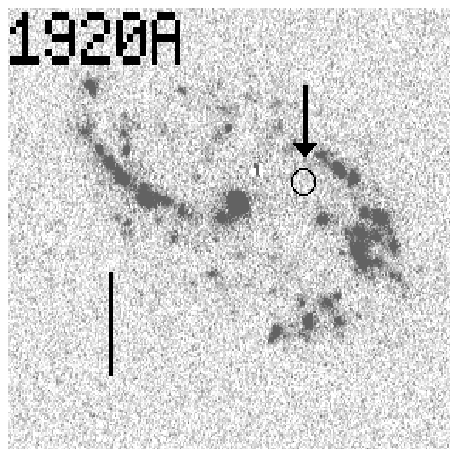}
\includegraphics[angle=0,width=4.5cm]{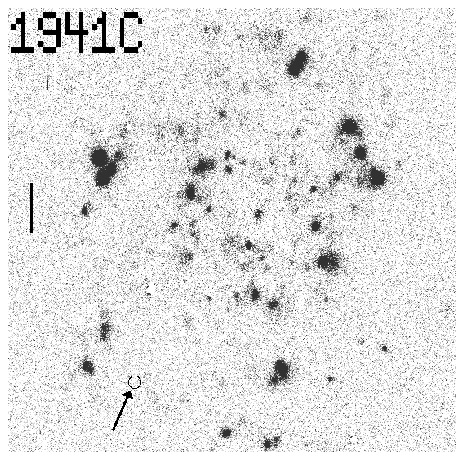}
\includegraphics[angle=0,width=4.5cm]{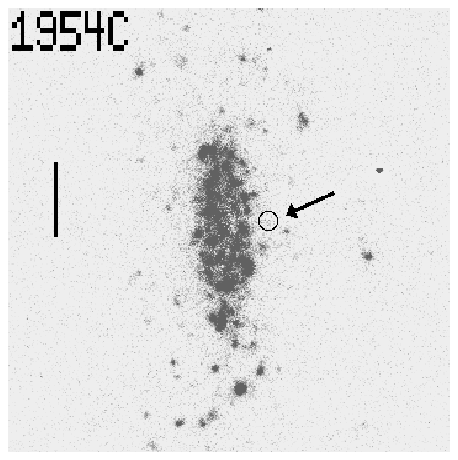}
\includegraphics[angle=0,width=4.5cm]{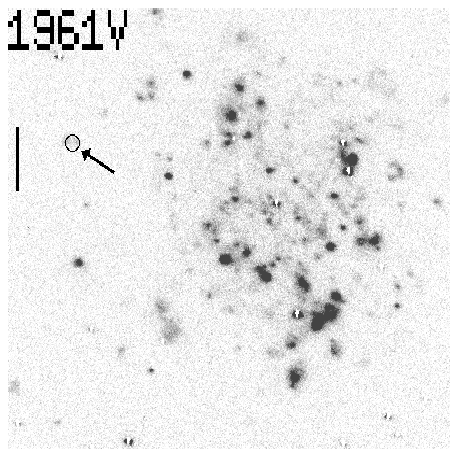}
\includegraphics[angle=0,width=4.5cm]{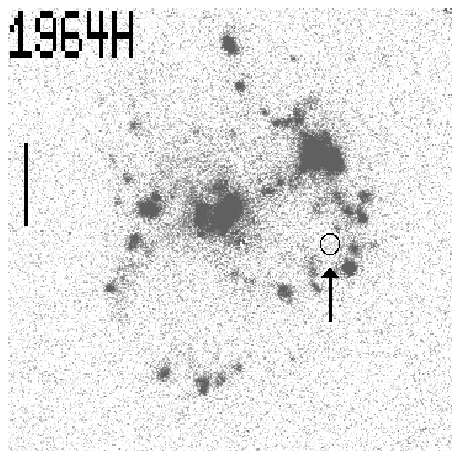}
\includegraphics[angle=0,width=4.5cm]{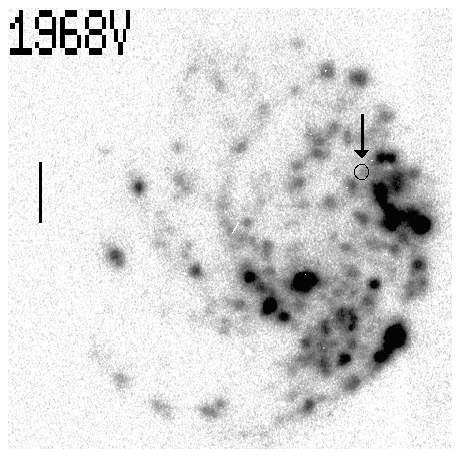}
\includegraphics[angle=0,width=4.5cm]{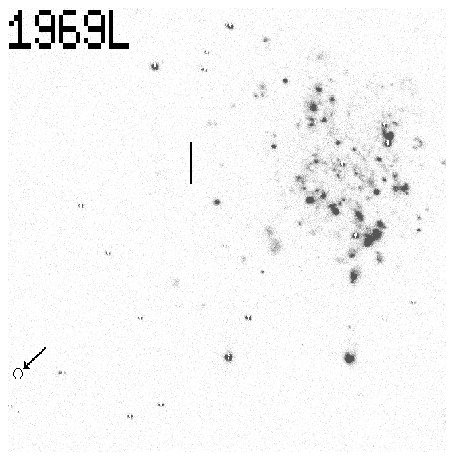}
\includegraphics[angle=0,width=4.5cm]{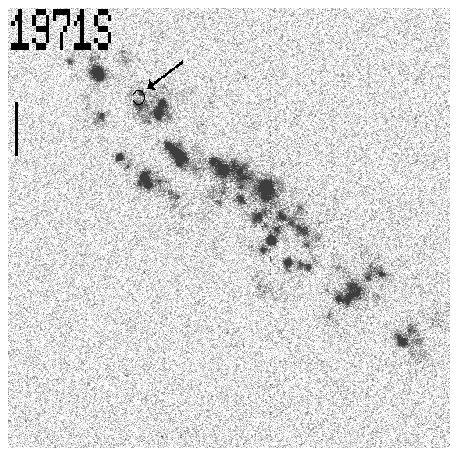}
\includegraphics[angle=0,width=4.5cm]{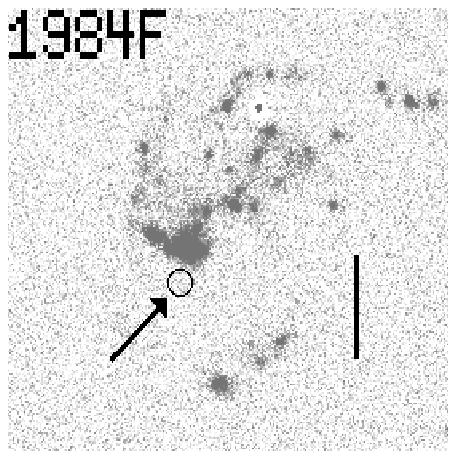}
\includegraphics[angle=0,width=4.5cm]{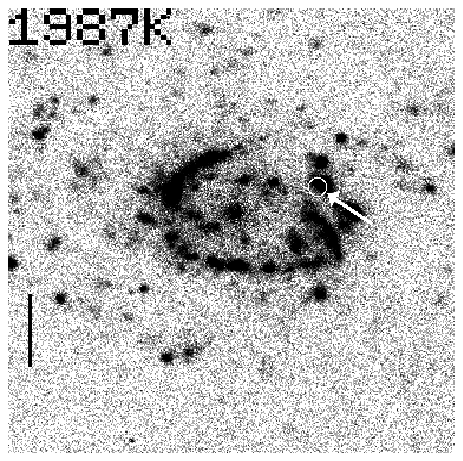}
\includegraphics[angle=0,width=4.5cm]{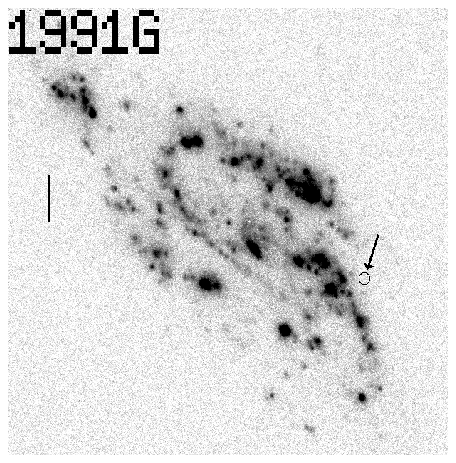}
\includegraphics[angle=0,width=4.5cm]{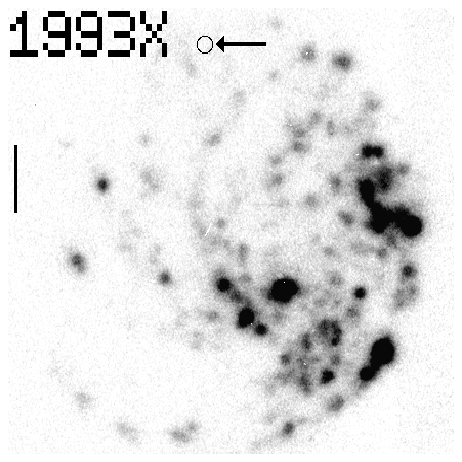}
\includegraphics[angle=0,width=4.5cm]{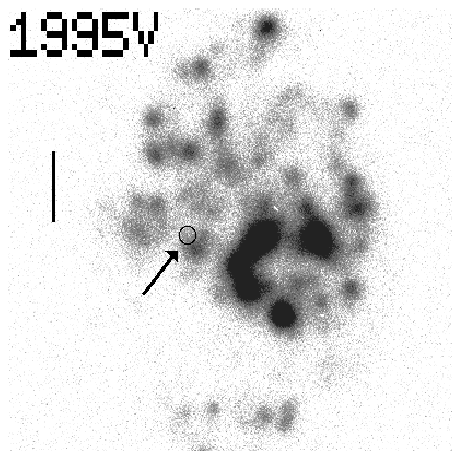}
\includegraphics[angle=0,width=4.5cm]{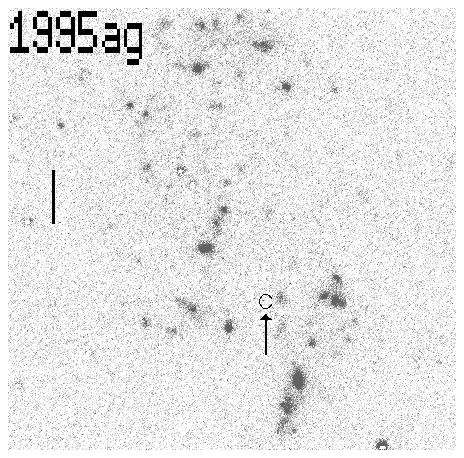}
\includegraphics[angle=0,width=4.5cm]{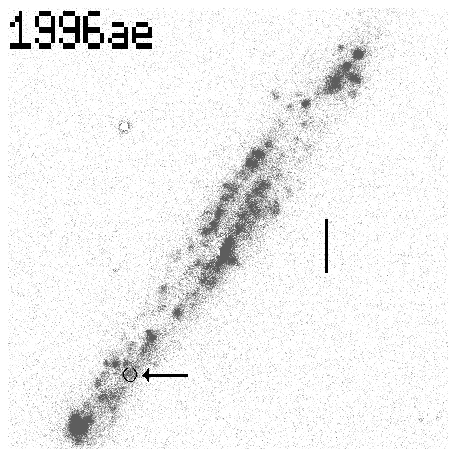}

\caption{Positions of Type II supernovae (circles) overlaid on negative images
showing the distribution of \Ha$+$\NII\ emission.  The scale bars are
20$^{\prime\prime}$ in length. (Continued over.)
}
\label{fig:haimagesi}
\end{figure*}

\begin{figure*}
\centering
\includegraphics[angle=0,width=4.5cm]{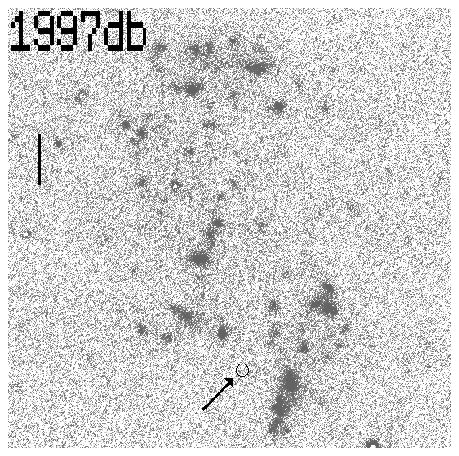}
\includegraphics[angle=0,width=4.5cm]{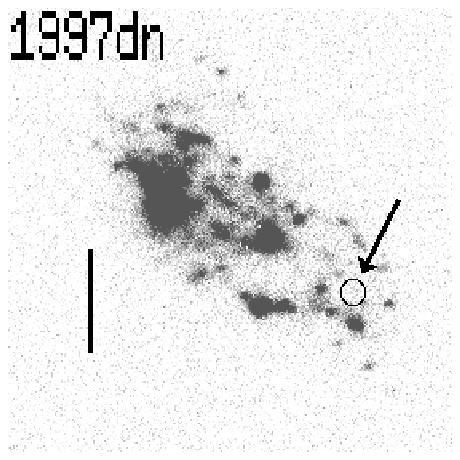}
\includegraphics[angle=0,width=4.5cm]{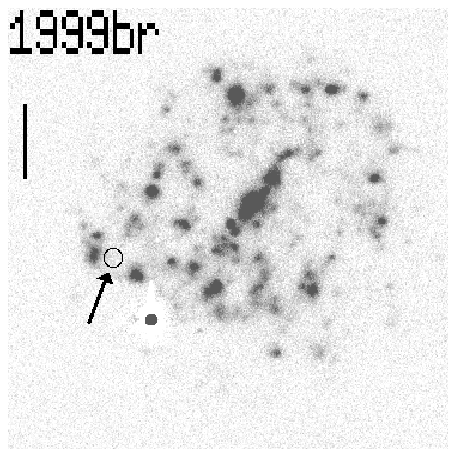}
\includegraphics[angle=0,width=4.5cm]{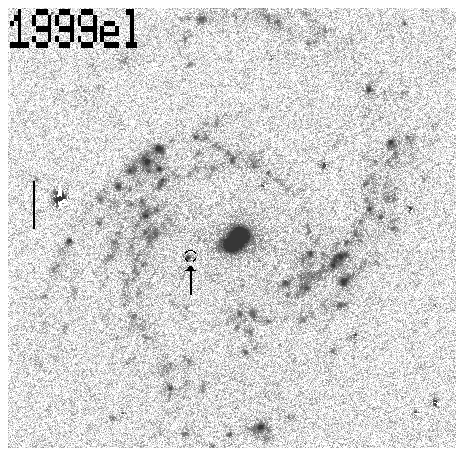}
\includegraphics[angle=0,width=4.5cm]{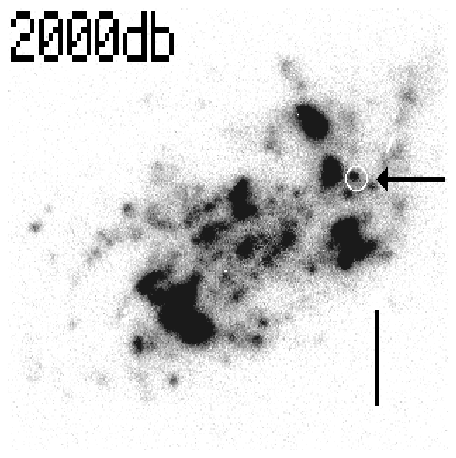}
\includegraphics[angle=0,width=4.5cm]{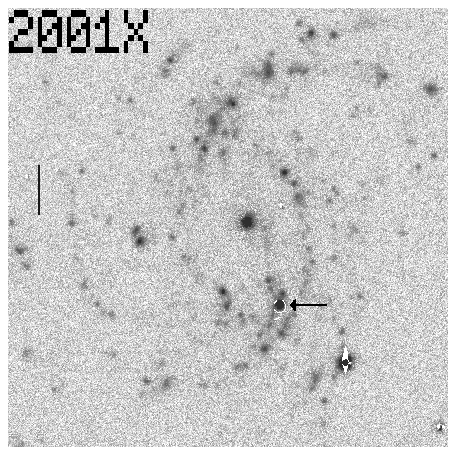}
\includegraphics[angle=0,width=4.5cm]{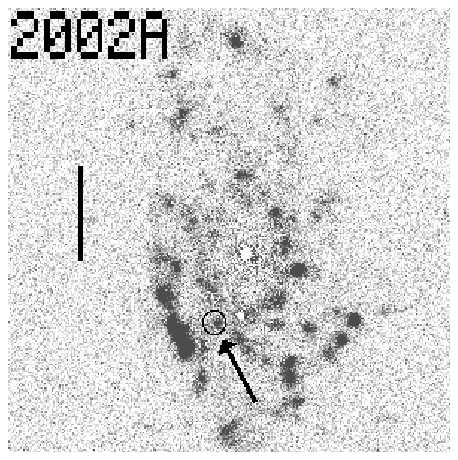}
\includegraphics[angle=0,width=4.5cm]{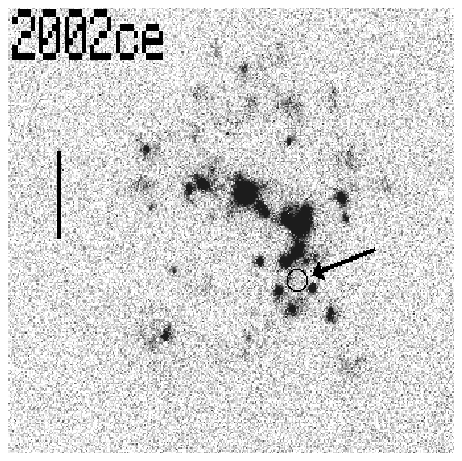}
\includegraphics[angle=0,width=4.5cm]{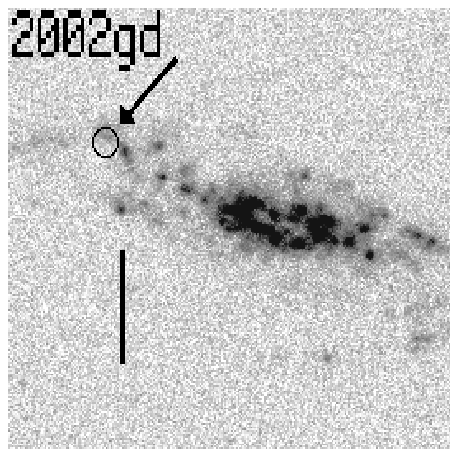}
\includegraphics[angle=0,width=4.5cm]{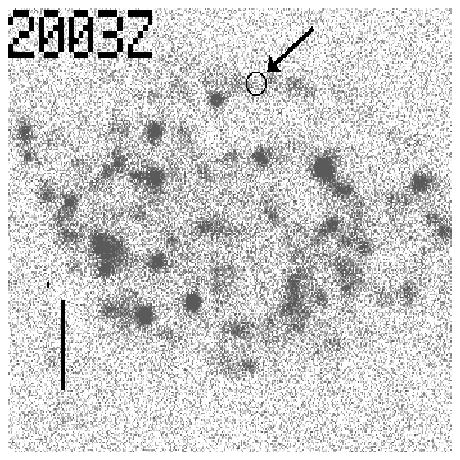}
\includegraphics[angle=0,width=4.5cm]{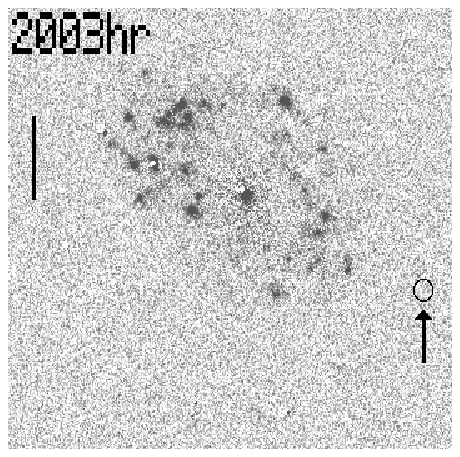}
\includegraphics[angle=0,width=4.5cm]{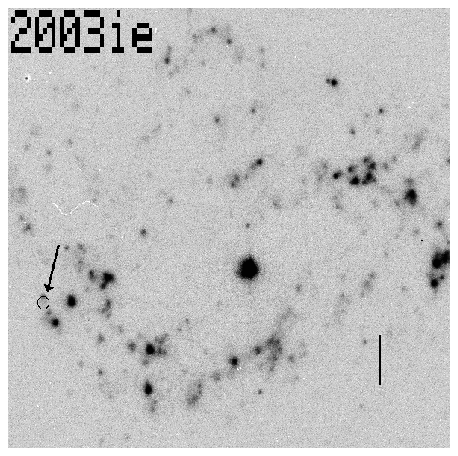}
\includegraphics[angle=0,width=4.5cm]{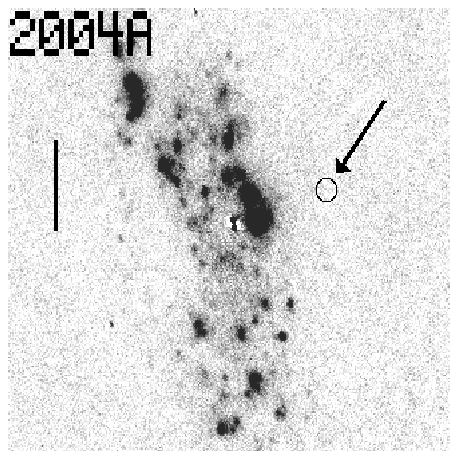}
\includegraphics[angle=0,width=4.5cm]{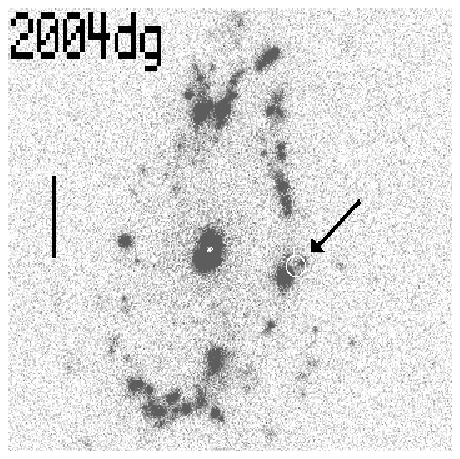}
\includegraphics[angle=0,width=4.5cm]{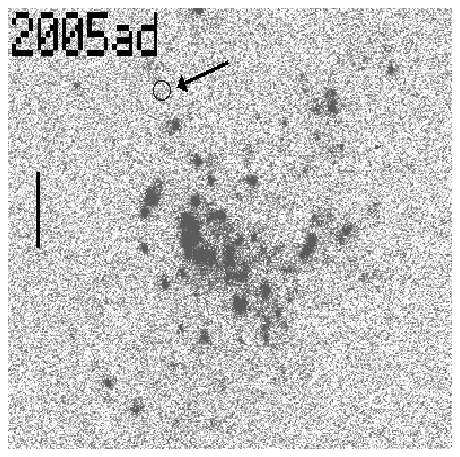}

\caption{Positions of Type II supernovae (circles) overlaid on negative images
showing the distribution of \Ha$+$\NII\ emission.  The scale bars are
20$^{\prime\prime}$ in length.
}
\label{fig:haimagesii}
\end{figure*}

\begin{figure*}
\centering
\includegraphics[angle=0,width=4.5cm]{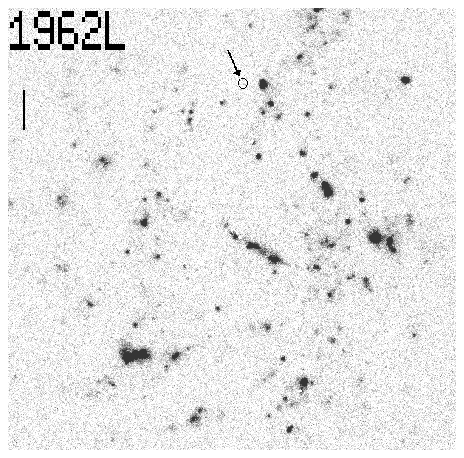}
\includegraphics[angle=0,width=4.5cm]{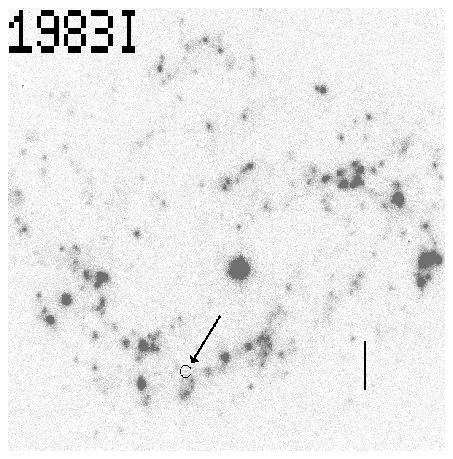}
\includegraphics[angle=0,width=4.5cm]{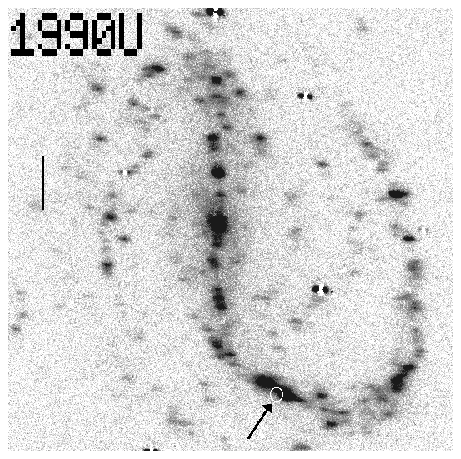}
\includegraphics[angle=0,width=4.5cm]{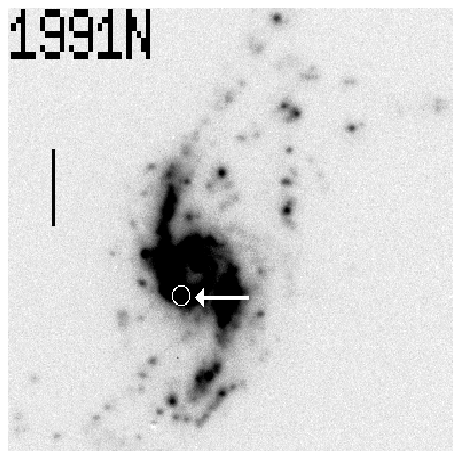}
\includegraphics[angle=0,width=4.5cm]{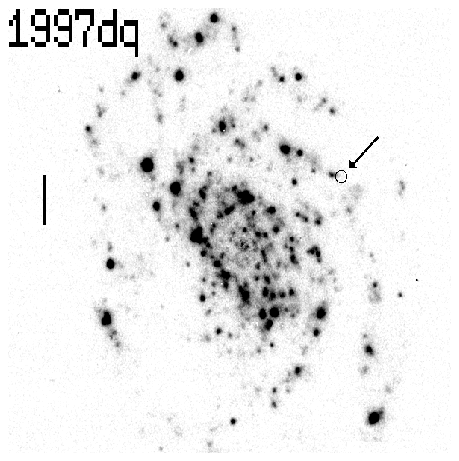}
\includegraphics[angle=0,width=4.5cm]{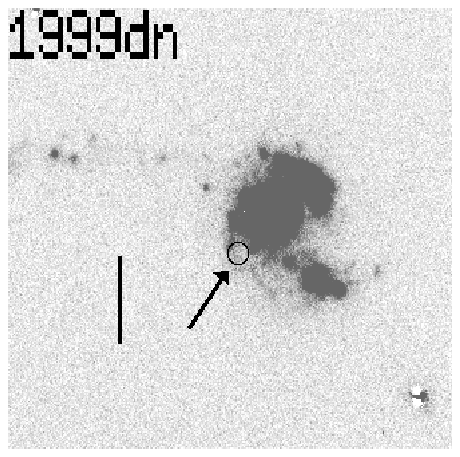}
\includegraphics[angle=0,width=4.5cm]{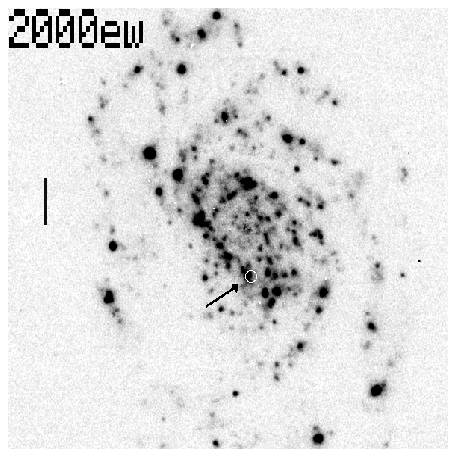}
\includegraphics[angle=0,width=4.5cm]{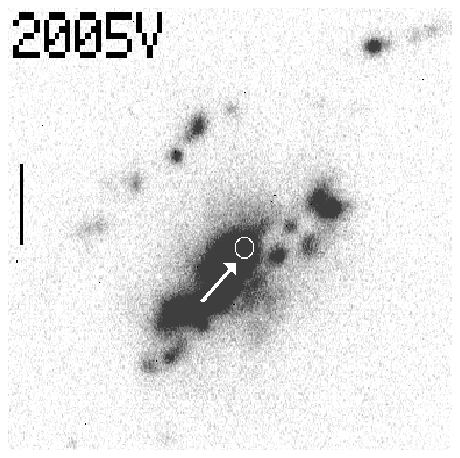}

\caption{Positions of Type Ib and Ic supernovae (circles) overlaid on negative images
showing the distribution of \Ha$+$\NII\ emission.  The scale bars are
20$^{\prime\prime}$ in length.
}
\label{fig:haimagesiii}
\end{figure*}

\begin{figure*}
\centering
\includegraphics[angle=0,width=4.5cm]{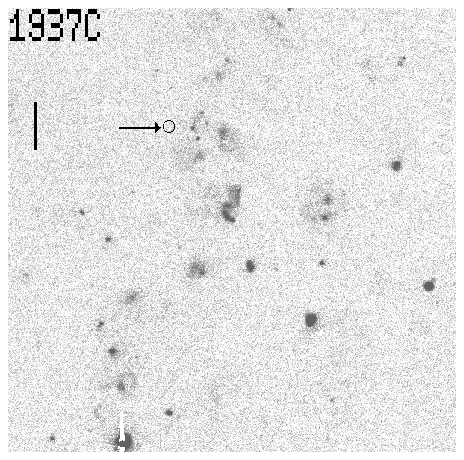}
\includegraphics[angle=0,width=4.5cm]{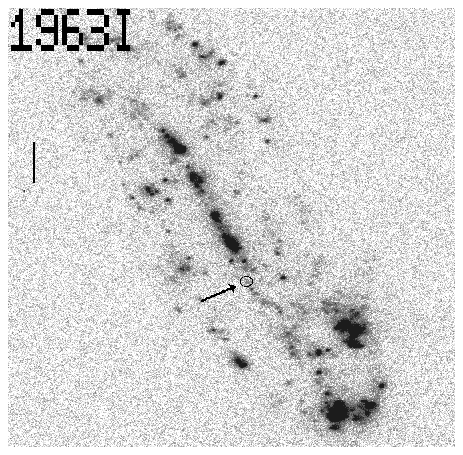}
\includegraphics[angle=0,width=4.5cm]{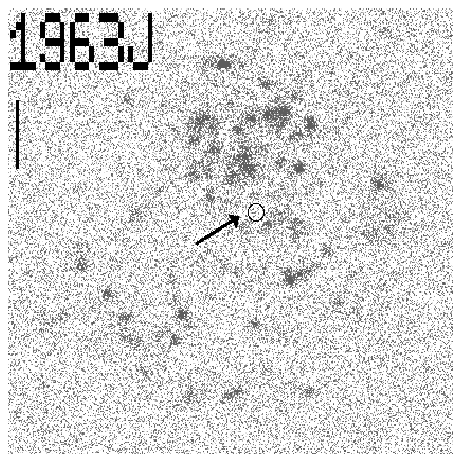}
\includegraphics[angle=0,width=4.5cm]{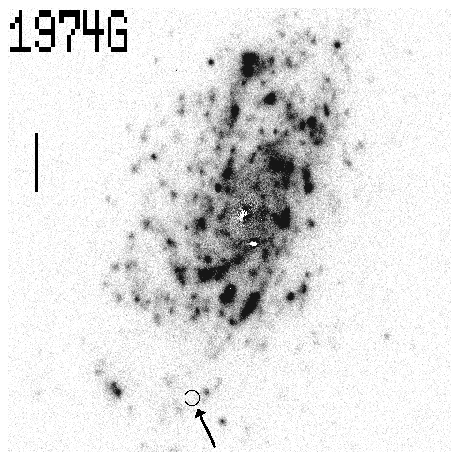}
\includegraphics[angle=0,width=4.5cm]{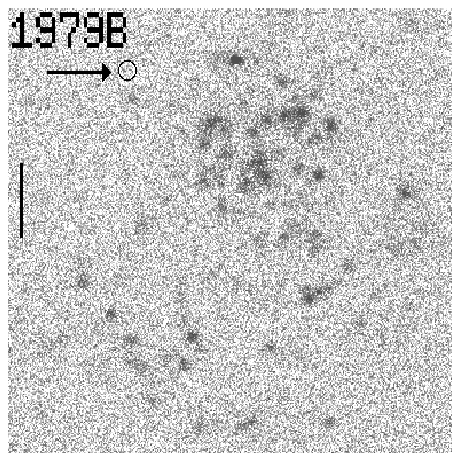}
\includegraphics[angle=0,width=4.5cm]{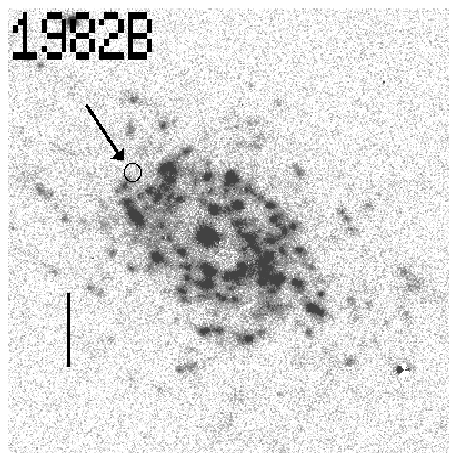}
\includegraphics[angle=0,width=4.5cm]{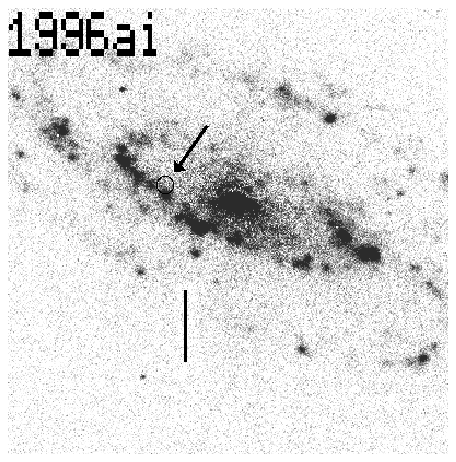}
\includegraphics[angle=0,width=4.5cm]{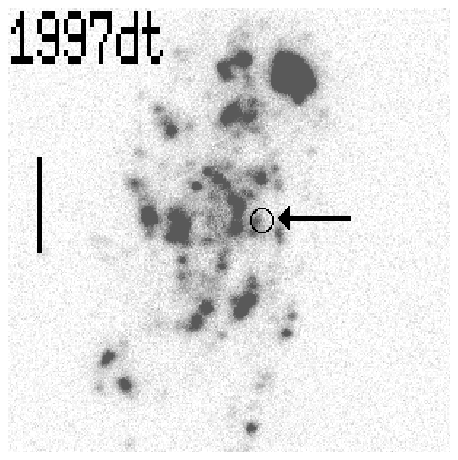}
\includegraphics[angle=0,width=4.5cm]{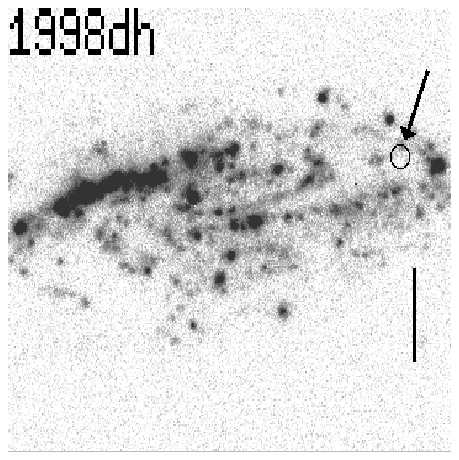}
\includegraphics[angle=0,width=4.5cm]{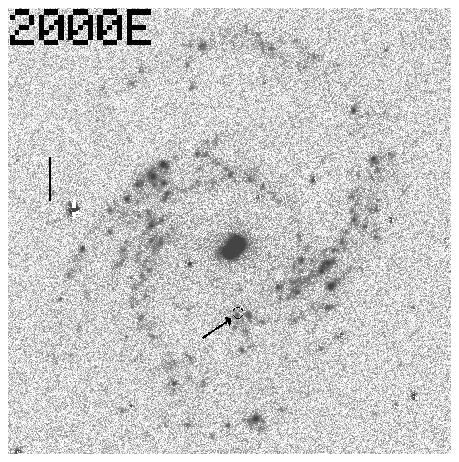}
\includegraphics[angle=0,width=4.5cm]{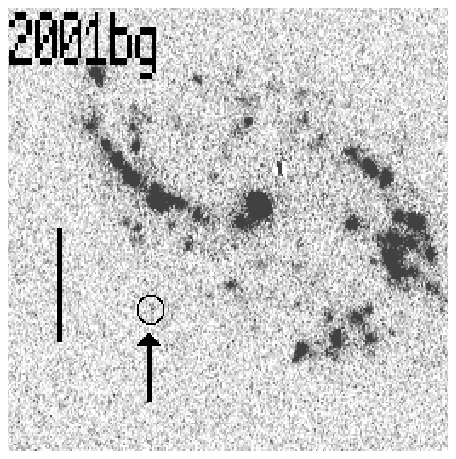}
\includegraphics[angle=0,width=4.5cm]{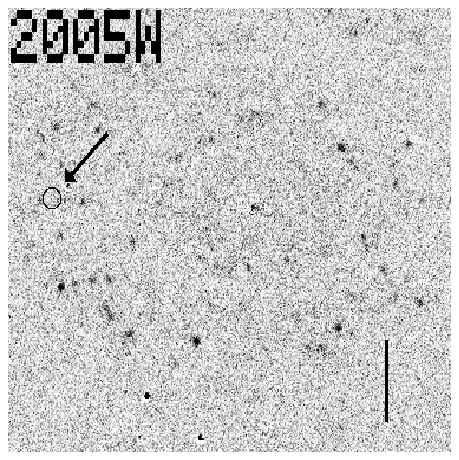}

\caption{Positions of Type Ia supernovae (circles) overlaid on negative images
showing the distribution of \Ha$+$\NII\ emission.  The scale bars are
20$^{\prime\prime}$ in length.
}
\label{fig:haimagesiv}
\end{figure*}

\begin{figure*}
\centering
\includegraphics[angle=0,width=4.5cm]{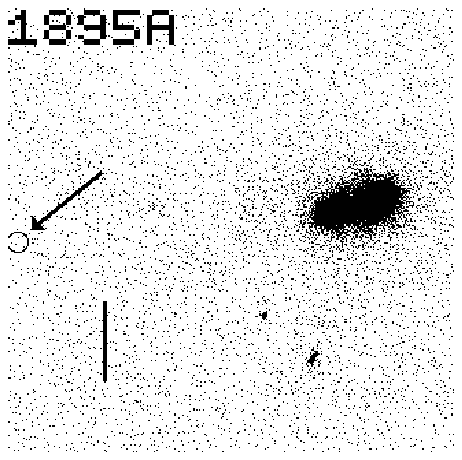}
\includegraphics[angle=0,width=4.5cm]{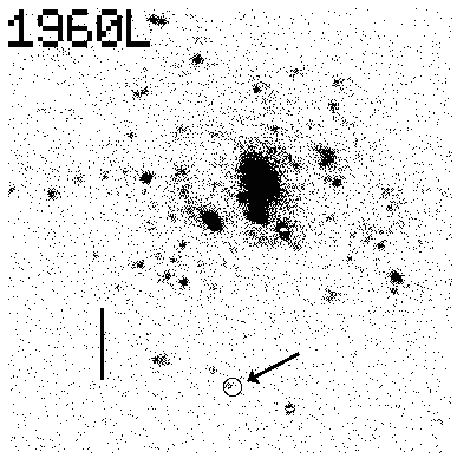}
\includegraphics[angle=0,width=4.5cm]{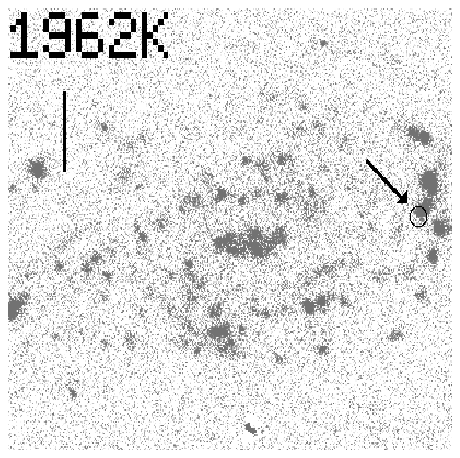}
\includegraphics[angle=0,width=4.5cm]{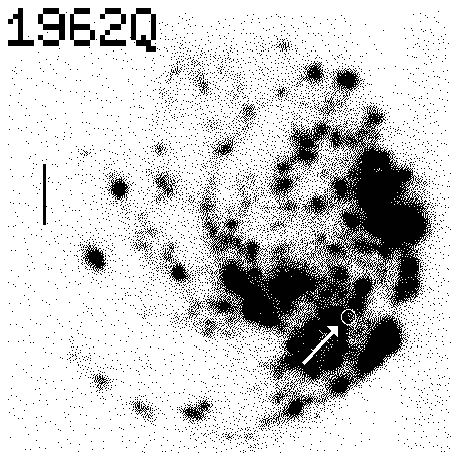}
\includegraphics[angle=0,width=4.5cm]{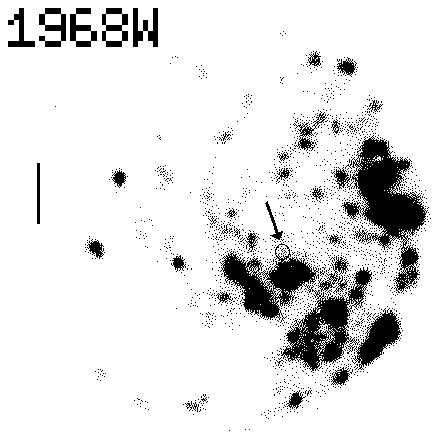}
\includegraphics[angle=0,width=4.5cm]{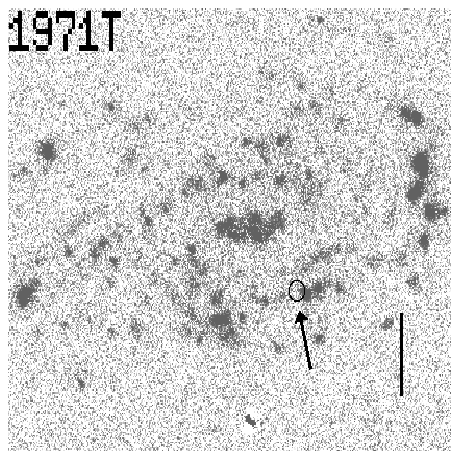}
\includegraphics[angle=0,width=4.5cm]{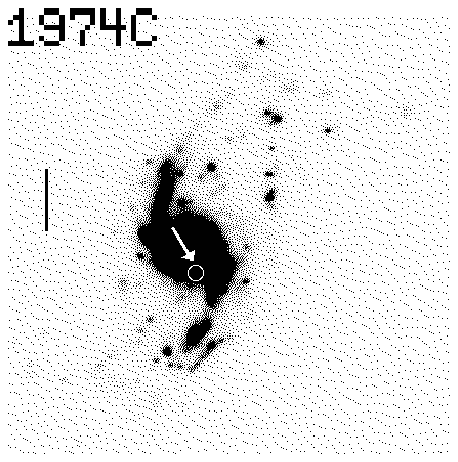}
\includegraphics[angle=0,width=4.5cm]{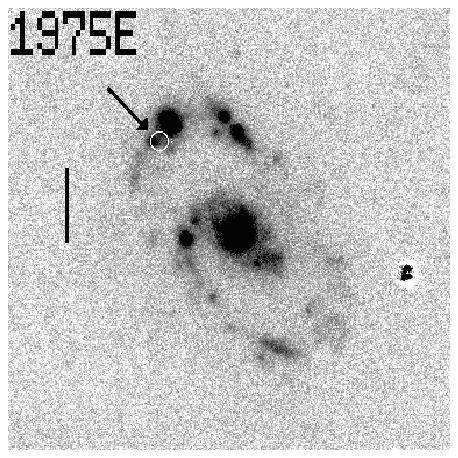}
\includegraphics[angle=0,width=4.5cm]{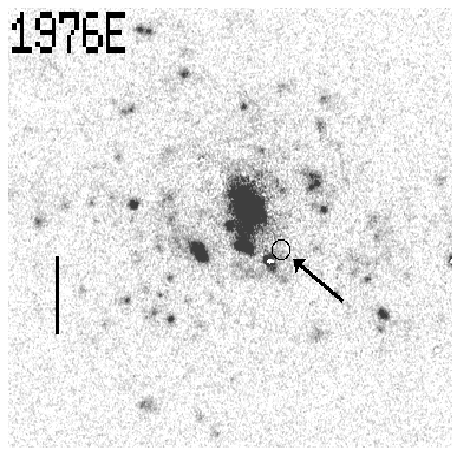}
\includegraphics[angle=0,width=4.5cm]{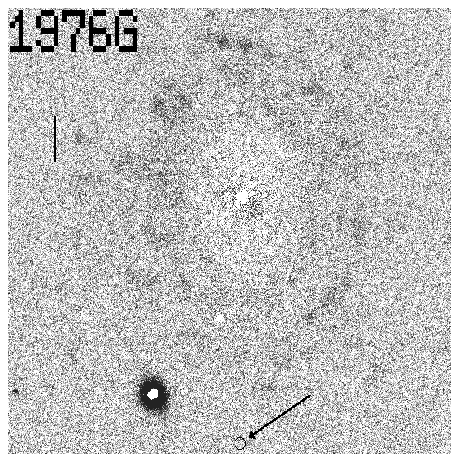}
\includegraphics[angle=0,width=4.5cm]{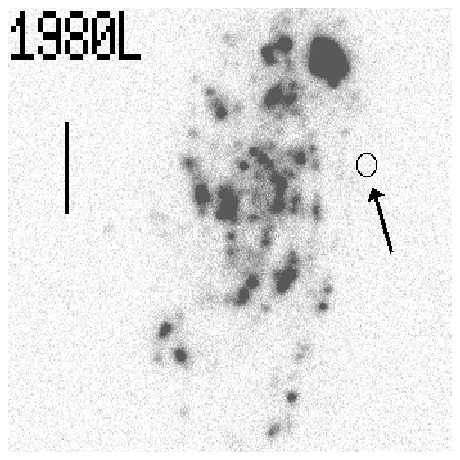}
\includegraphics[angle=0,width=4.5cm]{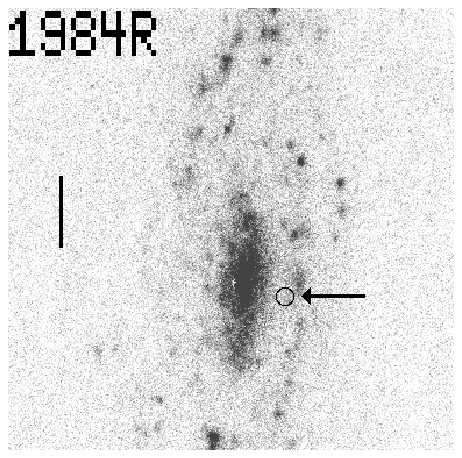}
\includegraphics[angle=0,width=4.5cm]{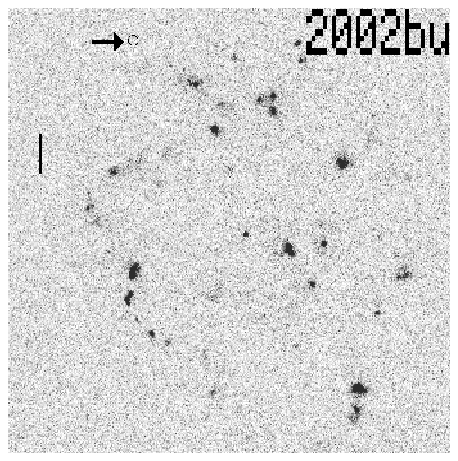}

\caption{Positions of unclassified supernovae (circles) overlaid on negative images
showing the distribution of \Ha$+$\NII\ emission.  The scale bars are
20$^{\prime\prime}$ in length.
}
\label{fig:haimagesv}
\end{figure*}

The \Ha GS image database (http://www.astro.livjm.ac.uk/HaGS/)
contains continuum-subtracted \Ha$+$\NII\ and $R$-band images for all
334 galaxies in the survey.  The pair of images for each galaxy is
fully reduced and accurately aligned at the sub-pixel level, as the
$R$-band images have generally been used for continuum subtraction of
the \Ha$+$\NII\ images, but none of them is astrometrically
calibrated.  We derived such calibrations for the 50 pairs of galaxy
images used in the present study in the following way.  The Canadian
Astronomy Data Centre website
(http://cadcwww.dao.nrc.ca/cadcbin/getdss) was used to download XDSS
second generation Palomar Sky Survey red images, matched in size
(10$^{\prime} \times$10$^{\prime}$)and position to each of the \Ha GS
image pairs.  It was found possible to match between 8 and 18 stars
between the XDSS and \Ha GS $R$ images, and use these to transfer the
accurate astrometry of the former to the latter.  The {\it Starlink}
GAIA image display tool was used to extract centroided Right Ascension
and Declination coordinates for each star in the XDSS image, and
corresponding centroided $(x,y)$ coordinates from the \Ha GS $R$ image
(stars are generally not visible on the \Ha$+$\NII\ images since these
are continuum-subtracted).  The {\it Starlink} ASTROM package was then
used to solve for a general conversion of $(x,y)$ coordinates to
(RA,Dec) values for each frame, with typical fit residuals of
0{\farcs}2 in each axis.  The ASTROM package was then used to convert
the (RA,Dec) coordinates of each SN into a $(x,y)$ coordinate, and the
resulting position is indicated by an arrowed circle in each \Ha$+$\NII\ image in Figs.
\ref{fig:haimagesi}--\ref{fig:haimagesv}. Figures \ref{fig:haimagesi}
and \ref{fig:haimagesii} show the galaxies hosting SNII;
Fig. \ref{fig:haimagesiii} those hosting SNIbc; Fig.
\ref{fig:haimagesiv} those hosting SNIa; and Fig. \ref{fig:haimagesv}
those hosting unclassified SNe.

Images are available in the literature for the following 14 supernovae
of the 63 in present study: 1937C \citep{baad38}, 1961V
\citep{zwic64}, 1962K \citep{zwic63}, 1964H \citep{zwic65}, 1974G
\citep{patc76}, 1982B \citep{ciat88}, 1987K \citep{fili88}, 1991G
\citep{blan95}, 1996ai \citep{ries99}, 1997dt \& 1998dh (Jha et al. 2005, 
astro-ph/0509234), 1999el \citep{dica02}, 2000E
\citep{vale03} and 2000ew \citep{vand03}.  These images range in
quality, but most are good enough to locate the supernova relative to
other prominent features in the galaxy at the $\sim$2$^{\prime\prime}$
level.  In every case our derived position was completely consistent
with the actual location of the supernova as shown in these images.

Relevant data for all SNe and host galaxies are listed in Table
\ref{tbl:1}.  Column 1 gives the supernova name; columns 2 and 3 the
NGC and UGC numbers of the host galaxies; column 4 the supernova type,
where known; columns 5 and 6 the host galaxy total star formation rate
(in solar masses/year) and log total $R$-band luminosities (in solar
units) respectively, both being taken from \Ha GS paper I; column 7
the location of the SN-containing image pixel in the normalised
cumulative ranked pixel value function (explained in section
\ref{sec:pixdist}); column 8 the fraction of the total galaxy
\Ha$+$\NII\ flux lying closer to the galaxy centre than the supernova
location (see section \ref{sec:raddist}); and column 9 the fraction of
the total galaxy $R$-band light lying closer to the galaxy centre than
the supernova location.


\section{An analysis of the SN - star formation spatial correlation 
based on image pixel statistics}
\label{sec:pixdist}

\subsection{Methods}

\begin{figure*}
\centering
\includegraphics[angle=-90,width=9.0cm]{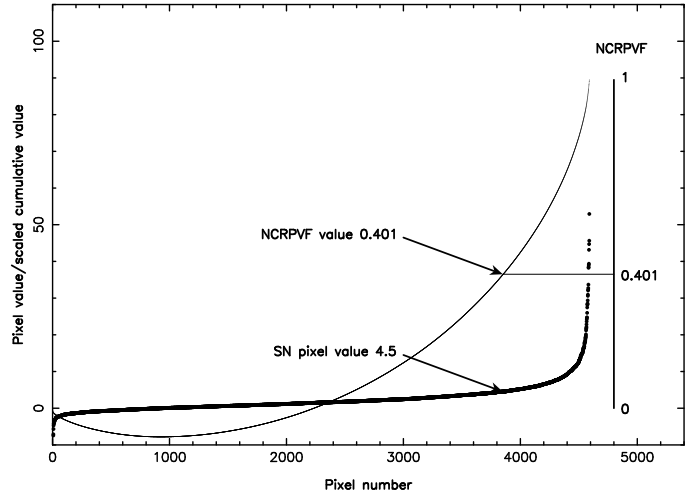}

\caption{Illustrative example showing the calculation of the NCRPVF index
for SN2002A in UGC~3804.  The larger points/thicker curved line shows the ranked
pixel values plotted against number in the ranked pixel sequence; the smaller 
points/thinner line shows the cumulative counts for each pixel, vertically scaled to 
fit. The pixel containing the location of SN2002A has a value of 4.5 counts, 
corresponding to an NCRPVF index of 0.401 as shown.   }

\label{fig:ncrpvf_ex}
\end{figure*}

Figures \ref{fig:haimagesi} and \ref{fig:haimagesii} show several
examples where SNII are apparently strongly associated with regions of
vigorous current star formation, as would be expected given the
high-mass progenitors of these SNe.  Examples are SN1987K, SN1999el,
SN2001X, SN2002A and SN2004dg.  However, there are also several
apparent counter-examples, e.g. SN1941C, SN1969L, SN2003hr and
SN2004A, where a SNII has occurred surprisingly far from any obvious
regions of ongoing star formation, as revealed by the \Ha$+$\NII\ line
emission.  Previous studies \citep[e.g.][]{bart94,vand96} have made
extensive analyses of this degree of correlation or non-correlation in
terms of the projected separation between each SN and the nearest
prominent \HII\ region.  It is clear from Fig. \ref{fig:haimagesi},
however, that the morphology of star formation in galaxies does not
always lend itself to the unambiguous definition of discrete regions
of star formation, and many star formation regions are faint or of low
surface brightness (e.g. that underlying the position of SN1961V in
Fig. \ref{fig:haimagesi}).  Some fraction of the ongoing star
formation must be taking place in low luminosity \HII\ regions and
hence it is possible that the apparently `unassociated' SNe could be
resulting from this population. We have thus formulated the following
test, based very loosely on the $V/V_{max}$ test used widely in other
areas of astrophysics, which addresses this specific question: Do SNe
of different types {\em quantitatively} trace the same stellar
population that gives rise to \Ha$+$\NII\ line emission? We test this
by analysing the pixel statistics of the sky- and
continuum-light-subtracted narrow-band images shown in
Figs. \ref{fig:haimagesi}--\ref{fig:haimagesv}.  The images are first
trimmed to the minimum size necessary to include all the line
emission, plus the location of the SN, to minimise the effects of sky
background uncertainty in the subsequent analysis.  We then bin the
pixels 3$\times$3, to reduce the pixel-to-pixel noise level and to
enable us to identify with a reasonable degree of certainty the pixel
containing the location of each supernova. This binning will typically
result in an array of 200$\times$200 pixel values.  The next stage is
to sort these pixel values in order of increasing pixel count, giving
a ranked sequence which starts with the most negative sky pixels and
ends with the pixels from the centres of the highest surface
brightness star formation regions.  Alongside this ranked distribution
of individual pixel values we also form the corresponding cumulative
distribution, by summing the pixel counts in the ranked sequence,
giving a sequence which counts up to the total emission line flux from
the galaxy.  We finally divide this cumulative rank sequence by the
total emission line flux, and set the initial negative values of
this cumulative function to zero, giving a normalised cumulative rank
pixel value function (NCRPVF henceforth) running from 0 to 1, with one
entry for each pixel in the input array.  This is illustrated for one 
typical case, SN2002A in UGC~3804, in fig. \ref{fig:ncrpvf_ex}.  In this 
case, the 4500 ranked pixels have values from --7 to $+$52, and are 
plotted with larger points, which merge to give the thicker curved line
running across the lower part of this plot.  The cumulative pixel sum, scaled
to fit on the same plot, is shown by the smaller points/thinner curved
line; note how this reaches a minimum where the individual pixel values
reach zero. The line to the right of the plot indicates the NCRPVF scale,
which runs from where the cumulative function reaches zero (NCRPVF$=$0) up 
to its maximum (NCRPVF$=$1).
In this case, the SN lies on a pixel with a value of 4.5 counts; reading 
vertically upwards to the cumulative scale  and then horizontally to the NCRPVF
scale indicates how the value of 0.401 is determined for this SN.

In forming the NCRPVF it is always found that a large fraction of the
pixels together contribute nothing to the final flux total; these
are defined as background pixels and are those for which the NCRPVF is
set to zero.  It is important to note that there are approximately
equal numbers of positive and negative pixels contributing to this sky
background region, as the cumulative function reaches a minimum when
the zero value of individual pixel values is reached, and it then
requires an equal number of positive pixels to bring the cumulative
function back to zero.  Of the remaining pixels which do contribute
positive counts to the total, the majority lie close to but just
above this positive sky level, but by force of numbers can contribute
a significant fraction of the total flux.  If the cumulative flux
distribution truly represents the total star formation activity in
each galaxy, and SNe of a particular type are drawn at random from
this same star forming population, then SNe should be equally likely
to come from any part of the NCRPVF.  Thus, over a long period of
time, the many pixels contributing to the 0-0.1 section of the NCRPVF
contribute as much to the total star formation rate of the galaxy, and
hence should host as many SNe, as the few bright pixels comprising the
0.9-1.0 section.  Thus, the test we perform is to see whether the
positions of the SN-containing pixels within the NCRPVF for their host
galaxy are uniformly distributed between 0 and 1.  Any population that
is randomly scattered throughout the host galaxies and not
preferentially associated with \HII\ regions will tend to lie at the
low end of the NCRPVF, since as stated above the huge majority of
pixels has low or zero values; any population purely associated with
the cores of bright \HII\ regions will tend to lie towards the high
end of the NCRPVF.

One obvious source of error in this method of calculating the NCRPVF
value is uncertainty in the precise position of the SN.  To quantify
this error, we repeated the calculation of the NCRPVF for each SN
location by replacing the exact pixel value at the SN location by the
median pixel value in the 3$\times$3 pixel box centred on that
location.  In almost all cases this made no significant difference to
the calculated NCRPVF value, and the mean value calculated using the
exact location was just 0.021 larger than when using the median pixel
of 9.  The RMS difference in NCRPVF value was 0.128, which was
dominated by two large differences of 0.555 and 0.660; with these
excluded the RMS difference fell to 0.070.  The two large differences
are for SN1937C and SN1995ag.  For both of these, the SN lies on a
fairly bright pixel surrounded by 8 which are consistent with sky
values.  Hence these may be due to cosmic rays which have escaped
detection, but since we have already binned the pixels up to
1$^{\prime\prime}$ these could be genuine small HII regions with most
of their emission line flux lying in one pixel.  Apart from these two
cases, the NCRPVF analysis seems to give results which are robust to
positional errors at the 1--2$^{\prime\prime}$ level.

A second possible source of error in the NCRPVF values is the adopted
sky level in the galaxy image.  This was checked by changing the sky
value to extreme maximum and minimum values and recalculating the
NCRPVF.  However, the changes found were small, as all the pixels in
the ranked distribution change together, and the only effect is on the
overall flux value used to normalise the position of the SN-containing
pixel in this ranked distribution.

We also performed a Monte-Carlo analysis of the effects of
pixel-to-pixel noise on the NCRPVF value.  This was done by adding a
further noise component to each pixel of the image used to derive the
NCRPVF value, including the SN-containing pixel.  The added noise had
a mean value of zero, and a standard deviation equal to that of a
blank sky region of the same image. Thus we were effectively
increasing the sky noise by a factor of 1.4 in each simulation.  This
was done several 10s of times for each image, with the NCRPVF value
being calculated each time.  The standard deviation of the simulated
NCRPVF values was then taken as the error on the NCRPVF value which we
quote in Table 1, i.e. the quoted error is a measure of how much
the position of the SN-containing pixel shifts in the normalised
distribution as a result of the extra noise being added to all pixels
in the image.  In general, the NCRPVF values were found to be quite
robust, in particular those from the extremes of the distribution.  A
few cases were found (SN1937C, SN1962K, SN1995ag and SN2002A; note
that two of these were also found to be sensitive to positional error
effects above) where this addition of noise can significantly affect
the derived value, but omission of these has no effect on the
conclusions presented here.  This also appears to be a random effect,
with no tendency to bias the results in one particular direction.

The NCRPVF analysis can also be applied to our $R$-band images of
these galaxies, but in this case the method produces very similar
results to the much simpler process of comparing supernova locations
with the radial growth curve of the galaxy light, which we discuss in
the next section.  This is because the brightest pixels in the
$R$-band always occur in the galaxy nucleus, and the ranked squence
thus runs monotonically from the outer regions of the galaxy into the
nucleus.  Thus, the $R$-band equivalents of figs. \ref{fig:ccpfrac}
and \ref{fig:Iapfrac} are essentially identical to
figs. \ref{fig:cchist}(b) and \ref{fig:snIahist}(b), which are
explained in section 5.

The value for each SN-containing pixel within its host
galaxy NCRPVF is listed in col. 7 of Table 1.  We will next discuss
the distributions of these values for the different types of SNe.
 
\subsection{Core-collapse supernovae}

\begin{figure*}
\centering
\includegraphics[angle=-90,width=7.0cm]{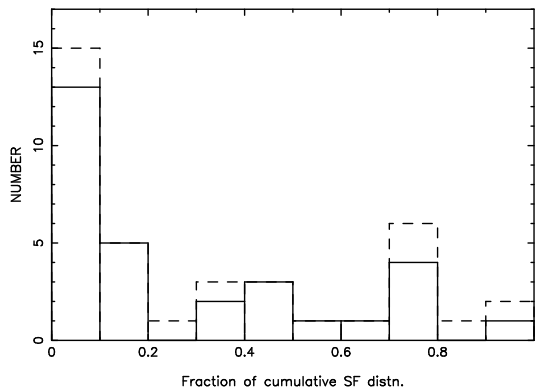}

\caption{Histograms of the positions of the SN-containing pixels
within the cumulative \Ha$+$\NII\ flux distributions for the 30 Type
II supernovae (solid line); and for all 38 core-collapse supernovae
(dashed line).  }

\label{fig:ccpfrac}
\end{figure*}

Figure \ref{fig:ccpfrac} shows the distribution of NCRPVF values for
the 30 SNII in the current sample (solid histogram), and the dashed
line indicates the distribution of all core-collapse SNe,
i.e. including the 8 SNIb, Ic and Ib/c (SNIbc henceforth).  It is
immediately clear that neither of these distributions is flat, with
the majority of these supernovae having low NCRPVF values,
corresponding to regions of no or low surface brightness line
emission.  The mean NCRPVF value for the 30 SNII is 0.274 with a
standard error on the mean of 0.053, while for all 38 core-collapse
SNe the mean is 0.320, standard error 0.051. These distributions are
confirmed by a Kolmogorov-Smirnov (KS) test to be inconsistent with
a flat distribution, at the $<1\%$ level. The excess is
concentrated in the two bins below 0.2, with 18 of the 30 SNII having
values below this limit, compared with an expectation value of 6 if
the SNII came from a parent population traced by the line emission.
Thus the excess of low NCRPVF values is 12, or 40\% of the total.  For
the 8 SNIbc considered alone, the conclusion is very different, with
these having a mean NCRPVF value of 0.490, standard error 0.121, and
being formally consistent with having been drawn from a flat parent
distribution, as confirmed by a KS test.  Hence the latter types are
consistent with being distributed evenly over the NCRPVF range of
0--1, and these SNe do seem to follow the same spatial distribution as
the line emission in their host galaxies, although the uncertainties
are great as a result of the small sample size.  It is noteworthy that
2 of the 3 highest NCRPVF values, and 4 of the 9 highest values, are
found for SNIbc, which only comprise 8 out of 63 SNe in the full
sample. Thus the SNIbc seem more likely than SNe of any other type to
lie close to the centres of luminous \HII\ regions, as would be
expected for SNe with the highest-mass progenitor stars.

\subsection{SNIa}

\begin{figure*}
\centering
\includegraphics[angle=-90,width=7.0cm]{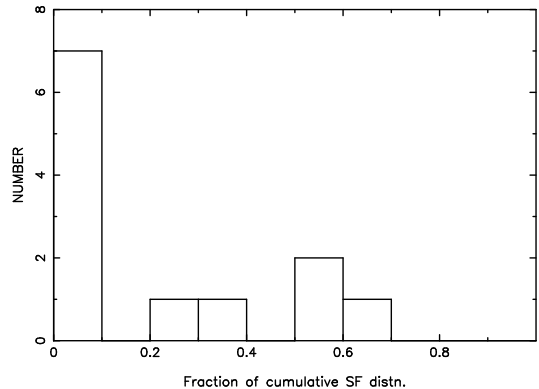}

\caption{Histogram of the positions of the SN-containing pixels within
the cumulative \Ha$+$\NII\ flux distributions for the 12 Type Ia
supernovae (solid line).  }

\label{fig:Iapfrac}
\end{figure*}

Figure \ref{fig:Iapfrac} shows the distribution of NCRPVF values for
the 12 SNIa.  As expected, this distribution is strongly weighted
towards low values, with a mean value of 0.188 and a standard error on
the mean of 0.069.  Again, a KS test shows $<$1\% chance of these
values having been drawn from a flat parent distribution, and indeed
the SNIa appear even more weakly correlated with line emission than
the SNII. It is natural to ask whether there is any correlation at all
between the locations of SNIa and the sites of active star formation,
or whether the distribution of values in Fig. \ref{fig:Iapfrac} is
consistent with that expected for random locations within the host
galaxies.  A simple test was performed to address this question, by
comparing the SN-containing pixel value with the median pixel value
over the subset image used for the NCRPVF calculation.  Clearly,
pixels chosen completely at random from these images would be equally
likely to lie above or below this median value, by definition.  It was
found the 8 of the SNIa locations had pixel values above the median,
and 4 below, showing a weak correlation between SNIa and ongoing star
formation.  It can also be argued, however, that this correlation
could reflect the concentration of all stellar populations towards the
centres of host galaxies, and may not imply any causal link.

\subsection{The full sample, and unclassified SNe}

We also looked at the distribution of NCRPVF values for all 63 SNe in
the current sample, and for the 13 unclassified SNe.  Again a strong
tendency was found for most NCRPVF values to be low, with a mean value
of 0.285, standard error 0.036 for the full sample.  The unclassified
SNe have a mean NCRPVF value of 0.272, standard error 0.063, and hence
are completely consistent with being drawn at random from the overall
population of SNe.  This indicates that no strong biases should be
caused by the incompleteness of type information; whatever the types
of the unclassified SNe, their inclusion in Figs. \ref{fig:ccpfrac}
and \ref{fig:Iapfrac} would have little or no effect on our
conclusions.

%
%

\section{Study of the radial distribution of SNe locations, relative to SF and
$R$-band light in host galaxies}
\label{sec:raddist}

\subsection{Methods}

In this section we perform a statistical test related to that in the
previous section, but in this case specifically to determine whether
the {\em radial} locations of SNe within their host galaxies are
consistent with their being drawn at random from the star-forming
population traced by \Ha$+$\NII\ emission, or the older stellar
population traced by $R$-band emission.  The data for this test were
the \Ha$+$\NII\ and $R$-band growth curves which were determined for
all \Ha GS galaxies using methods described in Paper I.  Briefly,
these curves were derived using elliptical apertures with fixed
centres, ellipticities and position angles, with values for the ellipse
parameters being taken from the UGC.  For each SN, we then calculated
the size of the elliptical aperture of the same shape and orientation
as its host galaxy which just included the location of the SN, and
thus were immediately able to calculate the fraction of both
\Ha$+$\NII\ and $R$-band light lying inside and outside the SN
location, by reference to the equivalent points on the growth curves.
If, for example, the core-collapse SNe are equally likely to come from
any part of the star-forming population of their host galaxies, we
would expect the fractions of \Ha$+$\NII\ light within the SN
locations to be evenly distributed between 0 and 1, with an average of
0.5 (note that throughout this analysis we quote the fractions {\it
within} the SN location, i.e. closer to the galaxy nucleus).  This is
the test we present in this section, splitting the SNe into types II,
all CC and Ia in turn, and testing all three categories against both
the \Ha$+$\NII\ and $R$-band light distributions of their host
galaxies.  We also look briefly at the properties of the unclassified
SNe, and of the total sample of 63 SNe.

\subsection{Core-collapse supernovae}

\begin{figure*}
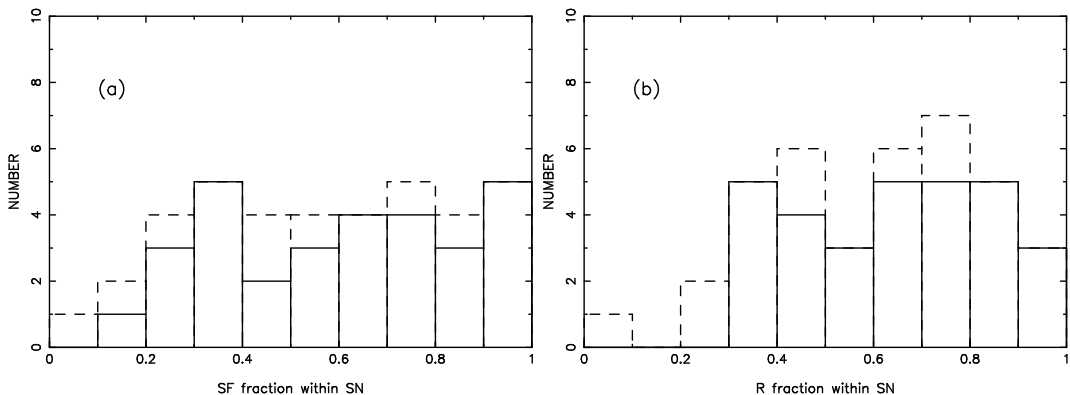

\centering
\includegraphics[angle=-90,width=7.0cm]{cchist.ps}
\includegraphics[angle=-90,width=7.0cm]{ccRhist.ps}

\caption{(a) Histograms of fractional \Ha$+$\NII\ fluxes
within the locations of 30 Type II supernovae (solid line); and all 38
core-collapse supernovae (dashed line).
(b) Histograms of fractional $R$-band fluxes
within the locations of 30 Type II supernovae (solid line); and all 38
core-collapse supernovae (dashed line).
}

\label{fig:cchist}
\end{figure*}

The solid line in Fig. \ref{fig:cchist}(a) shows the distributions of
fractions of \Ha$+$\NII\ emission coming from regions lying within the
locations of the SNe, for the host galaxies of the 30 SNII in the
present study.  The dashed line shows the same data, but with the
addition of the 8 SNIbc.  Overall, both histograms are close to the
flat distributions expected if these SNe perfectly trace the young
line-emitting stellar population.  The only deviation from a
completely flat distribution is the deficit in SNe arising from the
very central parts of the \Ha$+$\NII\ distribution.  Note that,
because the \Ha$+$\NII\ emission can exhibit significant central holes
e.g. in galaxies with purely old stellar bulges, a SN can in some
cases lie some distance from the physical centre of its host galaxy
and still be at the centre of the distribution of line emission in
Fig. \ref{fig:cchist}(a).  This departure from flat distributions was
shown by a KS test to be marginally significant, at the 0.10 level.
There is a slight tendency for the SNIbc to lie closer to the centres
of their host galaxies than the SNII, with the mean fractions of SF
lying within the SN locations being 0.612$\pm$0.045 for the SNe II,
against 0.446$\pm$0.088 for the SNIbc.  This difference is of marginal
statistical significance (1.7$\sigma$), but was also found by
\citet{vdb97} for a rather larger sample (16 SNIbc and 58 SNe II), and
it is noteworthy that, in the current study, the two lowest fractions
in Fig. \ref{fig:cchist}(a) correspond to SNIbc, SN2005V and
SN2000ew. This provides some support for the suggestion of metallicity
being linked to the difference between these SNII and SNIbc
\citep[discussed in][]{vdb97}.

Two explanations can be suggested for the apparent central deficit in
CC SNe.  The first and probably dominant effect is the systematic
failure to detect SNe in the central regions of galaxies due to either
the high surface brightness of the stellar background \citep{shaw79}
or extinction effects which have been deduced to result in significant
deficits in SN detection rates, even in near-IR searches
\citep{mann03}.  The second possible explanation is that there is an
additional emission-line component at the centres of these galaxies
which is not related to star formation, and hence not linked to CC
SNe.  Such a component could be related to AGN (although most of these
galaxies do not show strong unresolved central emission-line sources)
or to the extended nuclear emission-line sources noted by
\citet{hameed} and \citet{paper2}. If such a component were to lie
predominantly inside the innermost star formation regions, it could
contribute to the central drops in the histograms shown in Fig.
\ref{fig:cchist}.

We also note that the central deficit in SN numbers appears much
larger when the SNII distribution is compared with the $R$-band light
distribution (Fig. \ref{fig:cchist}(b)).  In this case, none of the
SNe lies within the central 30\% of the light distributions of their
host galaxies, and the difference from a flat distribution is now
quite significant, at the 0.01 level.  This greater deficit is
simply explained as being due to the central older bulge population
with no associated SNII.

\subsection{SNIa}

\begin{figure*}
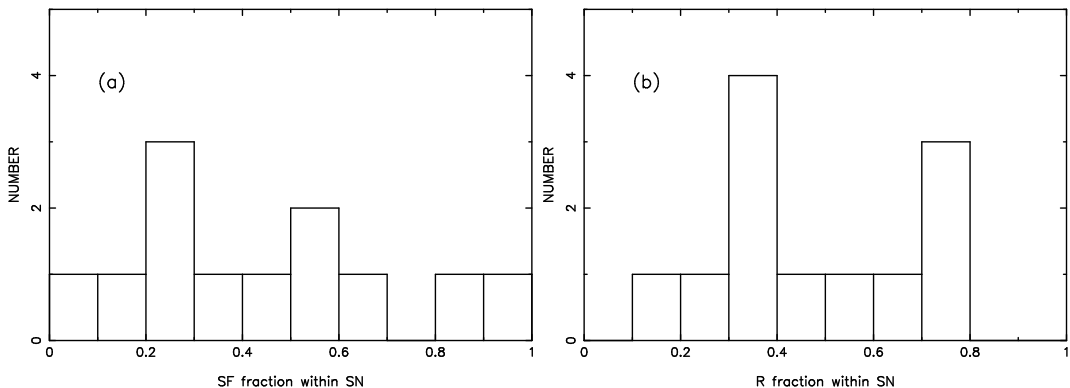

\centering
\includegraphics[angle=-90,width=7.0cm]{typeIahist.ps}
\includegraphics[angle=-90,width=7.0cm]{typeIaRhist.ps}

\caption{Histogram of (a) fractional \Ha$+$\NII\ fluxes, and (b) fractional
$R$-band fluxes, within the locations of the 12 Type Ia supernovae 
}
\label{fig:snIahist}
\end{figure*}

Figure \ref{fig:snIahist} shows the comparison of the positions of the
12 SNIa relative to the \Ha$+$\NII\ and $R$-band light distributions
of their host galaxies.  With the low-number statistics, both
histograms are formally consistent with flat distributions, and it is
impossible to determine from these data whether SNIa better trace the
young or old stellar populations.  The SNIa appear more
centrally-concentrated than the type II or CC SNe, with no evidence
for a central deficit in either Figs. \ref{fig:snIahist} (a) or (b).

\subsection{Unclassified supernovae}

We also looked briefly at the distributions of fractions of SF and of
$R$-band light within the locations of all 63 SNe in the present
study, and compared these with the values for the 13 unclassified SNe.
The central deficits noted for the CC SNe persist when the full
sample is considered, with mean fractions of 0.560$\pm$0.03 relative
to the SF distribution and 0.584$\pm$0.033 relative to the $R$-band
light distribution.  The unclassified SNe have distributions
consistent with their being drawn at random from the full distribution
of 63 SNe.

%
%

\section{Comparison of galactic supernova rates and host galaxy star formation rates}
\label{sec:snsfrcomp}

\subsection{Methods}

In this section, we perform a test of whether the
frequency of occurrence of SNe within galaxies appears to scale with
their total star formation rates, as revealed by \Ha$+$\NII\ fluxes,
or with their $R$-band luminosities.  One motivation for this test is
the suspicion that lower-luminosity galaxies may be less intensively
studied than their more spectacular counterparts, resulting in a
larger fraction of their SNe being missed entirely.  The test is in
some ways analogous to that performed with respect to light
distributions within individual galaxies, but here the test is
performed on the whole ensemble of \Ha GS galaxies, since the
probability of a detected supernova in any one of the galaxies is
small.  The only data used are the total \Ha$+$\NII\ fluxes (or
equivalently total star formation rates, where we use the conversion
formula of \cite{kenn98}) and total $R$-band luminosities of each of
the 327 \Ha GS UGC galaxies.  These total fluxes were sorted into
ranked sequences, running from the lowest intrinsic luminosity of the
327 up to the highest, and were then summed in that order to construct
cumulative luminosity distributions in both \Ha$+$\NII\ and $R$-band
light.  These were then normalised by the total luminosity in each band for all
327 galaxies, to give cumulative distributions running from 0 to 1. We
then tested to see where in these normalised cumulative distributions
lay the galaxies hosting the SNe of the different types already
discussed in this paper, and performed K-S tests to see whether they
were biased towards galaxies of high or low star formation rates or
luminosities.

Note that there is no assumption in this test that the \Ha GS sample in
any sense constitutes a statistically complete sample of
galaxies. Although it is representative of the star-forming galaxies in
the UGC, the magnitude and diameter selection criteria necessarily lead
to the very faintest and smallest galaxies being under-represented
relative to a truly volume-limited sample.  The question is simply
whether, for a given sample of galaxies selected in a way which is
hopefully unbiassed with respect to the likelihood of SNe occurring in
particular galaxies, the probability of a SN being detected in a
particular galaxy within that sample depends linearly on its current star
formation rate or $R$-band luminosity.

\subsection{Core collapse SNe}

\begin{figure*}
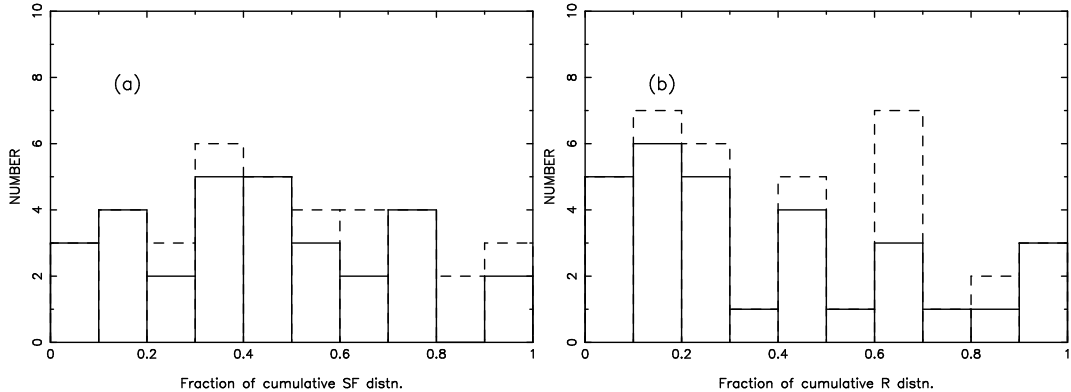

\centering
\includegraphics[angle=-90,width=7.0cm]{cc_cumSF.ps}
\includegraphics[angle=-90,width=7.0cm]{cc_cumR.ps}

\caption{(a) Histograms of the locations of galaxies in the
ranked cumulative \Ha$+$\NII\ distributions for 30 type II
supernovae (solid line) ; and all 38 core-collapse supernovae (dashed
line).
(b) Histograms of the locations of galaxies in the 
ranked cumulative $R$-band luminosity  distributions for 30 type II 
supernovae (solid line); and all 38 core-collapse supernovae (dashed 
line).
}
\label{fig:cumcchist}
\end{figure*}

As the first example of this test, the solid line in
Fig.\ref{fig:cumcchist} (a) shows the distribution of the locations of
the host galaxies of the 30 SNII in the ranked, cumulative \Ha$+$\NII\
distribution for the \Ha GS galaxy sample (solid line). The dashed
line shows the effect of adding in the 8 SNIbc to give the
distribution for all core-collapse SNe.  The distribution for SNII is
close to flat, but contrary to our expectations it shows a marginal
deficit in SNII within the galaxies with the highest star formation
rates (shown by the lack of points above 0.8 in the cumulative
distribution). The mean value for all 30 SNII is 0.435$\pm$0.047,
reflecting this marginally significant deficit at the high star
formation rate end.  A KS test gives a 20\% chance that these values
could have been drawn from a flat distribution.  The 8 SNIbc, on the
other hand, seem to come preferentially from galaxies with high star
formation rates, with a mean value of 0.651$\pm$0.087, higher than
that for the SNII with a significance of just over 2$\sigma$.  For all
core collapse SNe combined together, the distribution is statistically
completely flat, with a mean value of 0.481$\pm$0.044.  Thus it would
appear that galaxies with low star formation rates contribute at least
their expected numbers of core collapse SNe, and there is no evidence
for the expected selection bias against detecting SNe in such
galaxies.

Figure \ref{fig:cumcchist} (b) shows the test for the same SN host
galaxies, against the ranked cumulative distribution of $R$-band
luminosities of the 327 \Ha\ GS galaxies. Here again a tendency is
seen for the SNII (solid histogram) to come preferentially from the
fainter galaxies, with the mean location of the SNII-hosting galaxies
within the overall distribution being 0.367$\pm$0.053. A KS test gives
only a 5\% probability that these values are consistent with a flat
parent distribution.  The SNIbc values (the difference between the
dashed and solid histograms) are consistent with an unbiased
distribution with respect to galaxy $R$-band luminosity (mean
0.531$\pm$0.077), and the mean value for the galaxies hosting all 38
core-collapse SNe is 0.409$\pm$0.046.

\subsection{SNIa}

\begin{figure*}
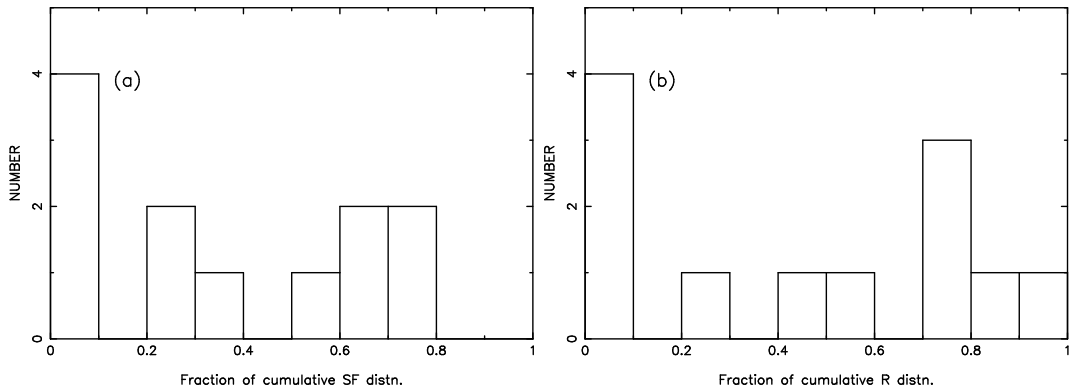

\centering
\includegraphics[angle=-90,width=7.0cm]{snIa_cumSF.ps}
\includegraphics[angle=-90,width=7.0cm]{snIa_cumR.ps}
\caption{Histogram of the locations of the galaxies hosting the 12
Type Ia supernovae in (a) the ranked cumulative \Ha$+$\NII\
distribution and (b) the ranked cumulative $R$-band light
distribution. }
\label{fig:cumIahist}
\end{figure*}

Figure \ref{fig:cumIahist} (a) shows the distribution of the
SNIa-hosting galaxies in the cumulative galaxy star formation
distribution, and displays a slight bias towards low values (mean
0.361$\pm$0.085, KS significance 0.20).  Figure \ref{fig:cumIahist}
(b) shows the equivalent values within the cumulative $R$-band light
distribution, which is formally consistent with a flat distribution
(mean 0.462$\pm$0.100).  Thus, as far as can be deduced from such
small numbers, the SNIa rates seem to scale with the $R$-band
luminosities of their host galaxies, and there is again no evidence of
any selection-induced deficit in SN numbers from low-luminosity
galaxies.  $R$-band luminosity appears a better predictor of SNIa
rates in these galaxies than star formation rate.

\subsection{All and unclassified SNe}

We repeated this analysis for the positions of all 63 SN-hosting
galaxies within the total star formation distribution and within the
total $R$-band light distribution.  Both tests show a small bias
towards low values (means 0.444$\pm$0.038 and 0.437$\pm$0.038
respectively), again showing that there is no evidence for selection
biases against low-luminosity galaxies.  The values for galaxies
hosting unclassified SNe are again completely consistent with being
drawn at random from the overall SN population analysed here.

The main conclusions from this section are that the SNIbc are the only
types which may preferentially occur in larger galaxies, and
particularly in those with high star formation rates, although this is
only of marginal significance due to the small number of SNIbc in our sample.
Other types of SNe show a slight tendency to occur preferentially in
fainter galaxies, and those with lower star formation rates.  This
second result also suffers from low number statistics, but if true it
is somewhat surprising.  One possible explanation is that SN detection
rates are higher in low luminosity galaxies due to smaller dust
extinction, although this explanation would require dust to have a
larger effect on SN rates than on emission of $R$-band and \Ha\
photons.  This extinction dependence is consistent with the interpretation
of SN detection rates in the near-IR in the study by \cite{mann03}.

\begin{figure*}
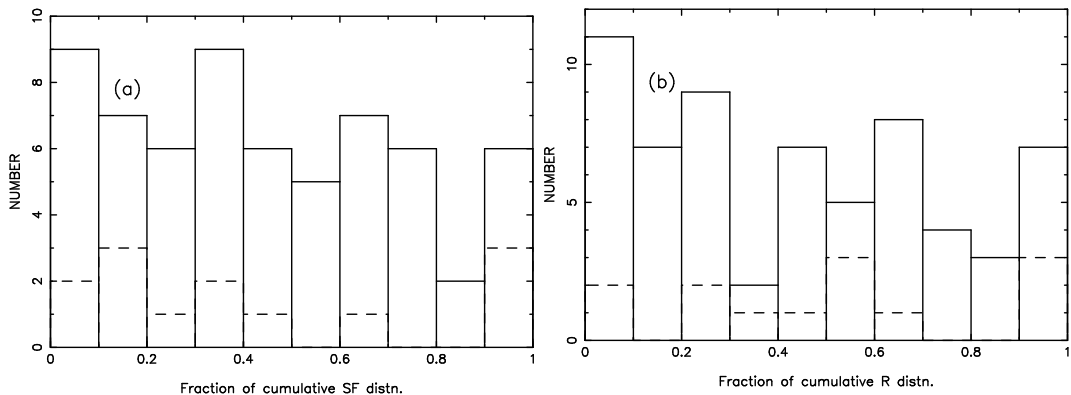

\centering
\includegraphics[angle=-90,width=7.0cm]{all_cumSF.ps}
\includegraphics[angle=-90,width=7.0cm]{all_cumR.ps}

\caption{(a) Histograms of the locations of galaxies in the
ranked cumulative \Ha$+$\NII\ distributions for all 63 supernovae in
the present study (solid line); and for the 13 unclassified supernovae
(dashed line).
(b) Histograms of the locations of galaxies in the 
ranked cumulative $R$-band luminosity  distributions for all 63
supernovae in the present study (solid line); and for the 13
unclassified supernovae 
(dashed line).
}
\label{fig:cumallhist}
\end{figure*}

\section{Discussion}

The most interesting result from this analysis is probably the finding
that many of the SNII in the current sample have no associated ongoing
star formation, as indicated by \Ha\ $+$ \NII\ line emission.  This
has also been found by previous authors, e.g. \citet{vand92},
\citet{vand96} and \citet{bart94}, and the significance of the present
result has been strengthened by the statistical analysis presented in
section \ref{sec:pixdist}.  This analysis shows that there is a real
excess, of $\sim$40\% of the total, of SNII in regions of zero or low
line flux, compared to what is expected if line emission accurately
traces high-mass star formation.  We will briefly discuss two
alternative (but not mutually exclusive) explanations for this result.

The first possibility is that each SNII progenitor did indeed form in
an HII region in the location indicated by the currently-measured line
emission, and then moved out of the HII region between birth and
death.  We estimate an upper limit for the time available for this
process from the lifetime of a 10~M$_{\odot}$ star, which current
stellar models put at 2.2$\times$10$^{7}$ years, independent of
metallicity to a good approximation.  As an illustrative calculation,
we have derived the component of velocity of the plane of the sky
required for some of the most anomalous SNII progenitors to have moved
to the location where the SN was observed, from the nearest bright HII
region.  We have done this for the 7 SNII with NCRPVF values of 0.000
in Table 1, i.e. those which appear to have no associated line
emission.  We use the galaxy distances listed in Paper I.  The
results are listed in Table \ref{tbl:velcalc}, where col. 1 gives the
SN name, col. 2 the host galaxy UGC number, col. 3 the galaxy distance
in Mpc, col. 4 the projected angular distance from the SN to the
centre of the nearest bright HII region, in seconds of arc, col. 5 the
same projected distance in kpc, and col. 6 the lower limit on the
estimated progenitor velocity.

\begin{table*}
\begin{center}
\begin{tabular}{llrrcc}
\hline
\hline
SN  & UGC & D(Mpc) & R$_p$($^{\prime\prime}$) & R$_p$(kpc) & V$_{\star}$(km~s$^{-1}$) \cr
\hline
2005ad &  1954 & 18.1 &  10.2 & 0.90 &  41\\
2004A  & 10521 & 15.0 &  10.9 & 0.79 &  36\\
2003hr &  4362 & 33.1 &  20.2 & 3.24 & 147\\
2003Z  &  4779 & 21.3 &   9.5 & 0.98 &  45\\
1969L  &  2193 &  6.9 & 102.0 & 3.41 & 155\\
1941C  &  7634 &  6.8 &  18.5 & 0.61 &  28\\
1920A  &  4484 & 32.1 &   7.2 & 1.12 &  51\\ 

\hline
\end{tabular}
\caption[]{Distance and drift velocity data for SNII with no associated line emission.}
\label{tbl:velcalc}
\end{center}
\end{table*}

The required velocity components of the progenitor stars are between
28 and 155~km~s$^{-1}$, with a median of 45~km~s$^{-1}$.  These values
are very comparable to those attributed to the OB runaway stars, which
are traditionally defined as having a minimum peculiar velocity of
40~km~s$^{-1}$ \citep{blaa61}, and may reach values as large as
200~km~s$^{-1}$ \citep{dray05}.  The fraction of OB stars that are
runaways has been estimated at between 7\% and 49\% \citep[and
refererences therein]{gies86}.  Thus the estimated 40\% excess of SNII
lying away from regions of present star formation could be
explained by the runaway phenomenon, but the fraction is uncomfortably
large and certainly lies at the upper end of the frequency of stars
previously thought to be
affected by this phenomenon.  It would also seem impossible to explain
such a high fraction of runaway stars through the binary-binary
encounter mechanism first proposed by \cite{pove67}, which ejects only
one of the four massive stars involved in the ejection mechanism,
whereas the binary supernova hypothesis of \cite{blaa61} could
presumably result in up to 50\% of massive stars being ejected from
the star formation region where they formed, and hence may be consistent with 
our findings.

An alternative explanation is that H$\alpha$ line emission is not a
good tracer of all star formation, and that $\sim$40\% of all star
formation takes place in locations from which little or no line
emission is produced, even though high mass stars are being produced.
This could occur if the star formation process is both rapid and
efficient, depleting the gas reservoir so that no detectable ionized
gas remains when SNe occur.  Alternatively, the dust geometry might be
such that the ionized gas is not visible, even though the SNe are visible. 

Due to the small number of SNIa in the present sample, it is hard
to draw any firm quantitative conclusions from these observations.  It
is, however, of interest to consider the environments of these 12 SNIa
in light of recent suggestions that there may be two populations of
SNIa, with `prompt' SNIa occurring shortly after the formation of
their parent stellar population, and `delayed' SNIa occurring when the
stellar population is typically some Gyr old \citep{mann05}.
The pixel analysis of the locations of SNIa relative to H$\alpha$
emission showed clearly and unsurprisingly that these SNe as a whole
do not trace the young stellar population in their host galaxy, and 7
of the 12 (58\%) have NCRPVF values consistent with coming from
background regions with no ongoing star formation.  These 7 are most
likely to be associated with the `delayed' fraction of SNIa.  The
remaining 5 (42\%) generally show good degrees of correlation with
sites of active star formation, and it is tempting to ascribe these to
the `prompt' fraction of SNIa.  However, it should be remembered that
some of these may be coincidental alignments of intrinsically old
systems with regions of the galaxy undergoing later bursts of star
formation, so the number of true `prompt' SNIa may be smaller than 5,
and comparison of figs. \ref{fig:Iapfrac} and \ref{fig:snIahist}(b) 
demonstrates that overall the SNIa
more closely follow the older stellar population traced by $R$-band
light than they do the extremely young stars traced by H$\alpha$.
Thus, with the current data we cannot exclude the possibility that all
SNIa come from an old stellar population, but our preferred conclusion
is that while the majority seem to be linked to the old stellar population,
a significant minority ($\leq$42\%) are linked to regions of ongoing
star formation.
 
\section{Conclusions}
\label{sec:conc}
\renewcommand{\labelitemi}{$\bullet$}

\begin{itemize}
\item From a pixel statistics analysis, we find an excess of SNII coming
from regions of no or weak star formation activity, as traced by \Ha\ line emission.
\item SNIbc do appear to trace star formation activity, with many of them coming 
from the centres of bright star formation regions.
\item SNIa are only marginally more likely to be found towards star formation 
regions than would be expected by chance, but our data are consistent with a 
significant minority, $\leq$42\%, being associated with ongoing star formation.
\item The radial distribution of CC SNe closely follows that of the \Ha\ line emission.
\item We find a lack of SNII from central regions of their host galaxies, possibly due 
to a component of the line emission not
associated with SF; alternative explanations include extinction effects, or SN missed
in high SB regions even in nearby galaxies.  
\item There is some evidence that SNIbc are more centrally concentrated than SNII, possibly 
indicating that the former preferentially occur in high metallicity regions
\item Overall, detected CC SN rates scale approximately linearly with both their host galaxy
luminosities and star formation rates.  However, relative to this overall trend, the 
SNII show a weak bias towards low-luminosity (and star formation rate) galaxies, and the 
SNIbc towards high-luminosity (and star formation rate) galaxies.
\item SNIa rates scale more closely with galaxy $R$-band luminosity than with star 
formation rates. 
\end{itemize}

\begin{acknowledgements}

The Jacobus Kapteyn Telescope was operated on the island of La Palma
by the Isaac Newton Group in the Spanish Observatorio del Roque de los
Muchachos of the Instituto de Astrof\'\i sica de Canarias.  This
research has made use of the NASA/IPAC Extragalactic Database (NED)
which is operated by the Jet Propulsion Laboratory, California
Institute of Technology, under contract with the National Aeronautics
and Space Administration.  The referee is thanked for many useful suggestions
which significantly improved the clarity of presentation of the paper.
PJ thanks Andy Newsam for assistance with
the analysis, and Ivan Baldry, Dan Brown and Maurizio Salaris 
for useful discussions.

\end{acknowledgements}
\bibliographystyle{bibtex/aa}
\bibliography{refs}

\begin{thebibliography}{37}
\expandafter\ifx\csname natexlab\endcsname\relax\def\natexlab#1{#1}\fi

\bibitem[{{Baade} \& {Zwicky}(1938)}]{baad38}
{Baade}, W. \& {Zwicky}, F. 1938, \apj, 88, 411

\bibitem[{{Bartunov} {et~al.}(1994){Bartunov}, {Tsvetkov}, \&
  {Filimonova}}]{bart94}
{Bartunov}, O.~S., {Tsvetkov}, D.~Y., \& {Filimonova}, I.~V. 1994, \pasp, 106,
  1276

\bibitem[{{Blaauw}(1961)}]{blaa61}
{Blaauw}, A. 1961, \bain, 15, 265

\bibitem[{{Blanton} {et~al.}(1995){Blanton}, {Schmidt}, {Kirshner}, {Ford},
  {Chromey}, \& {Herbst}}]{blan95}
{Blanton}, E.~L., {Schmidt}, B.~P., {Kirshner}, R.~P., {et~al.} 1995, \aj, 110,
  2868

\bibitem[{{Ciatti} {et~al.}(1988){Ciatti}, {Barbon}, {Cappellaro}, \&
  {Rosino}}]{ciat88}
{Ciatti}, F., {Barbon}, R., {Cappellaro}, E., \& {Rosino}, L. 1988, \aap, 202,
  15

\bibitem[{{Di Carlo} {et~al.}(2002){Di Carlo}, {Massi}, {Valentini}, {Di
  Paola}, {D'Alessio}, {Brocato}, {Guidubaldi}, {Dolci}, {Pedichini},
  {Speziali}, {Li Causi}, {Caratti o Garatti}, {Cappellaro}, {Turatto},
  {Arkharov}, {Gnedin}, {Larionov}, {Benetti}, {Pastorello}, {Aretxaga},
  {Chavushyan}, {Vega}, {Danziger}, \& {Tornamb{\' e}}}]{dica02}
{Di Carlo}, E., {Massi}, F., {Valentini}, G., {et~al.} 2002, \apj, 573, 144

\bibitem[{{Dray} {et~al.}(2005){Dray}, {Dale}, {Beer}, {Napiwotzki}, \&
  {King}}]{dray05}
{Dray}, L.~M., {Dale}, J.~E., {Beer}, M.~E., {Napiwotzki}, R., \& {King}, A.~R.
  2005, \mnras, 883

\bibitem[{{Filippenko}(1988)}]{fili88}
{Filippenko}, A.~V. 1988, \aj, 96, 1941

\bibitem[{{Gies} \& {Bolton}(1986)}]{gies86}
{Gies}, D.~R. \& {Bolton}, C.~T. 1986, \apjs, 61, 419

\bibitem[{{Greggio}(2005)}]{greg05}
{Greggio}, L. 2005, \aap, 441, 1055

\bibitem[{{Hameed} \& {Devereux}(1999)}]{hameed}
{Hameed}, S. \& {Devereux}, N. 1999, \aj, 118, 730

\bibitem[{{James} {et~al.}(2004){James}, {Shane}, {Beckman}, {Cardwell},
  {Collins}, {Etherton}, S., {Fathi}, {Knapen}, {Peletier}, {Percival},
  {Pollacco}, {Seigar}, \& {Steele}}]{paper1}
{James}, P.~A., {Shane}, N.~S., {Beckman}, J.~E., {et~al.} 2004, \aap, 414, 23

\bibitem[{{James} {et~al.}(2005){James}, {Shane}, {Knapen}, {Etherton}, \&
  {Percival}}]{paper2}
{James}, P.~A., {Shane}, N.~S., {Knapen}, J.~H., {Etherton}, J., \& {Percival},
  S.~M. 2005, \aap, 429, 851

\bibitem[{{Johnson} \& {MacLeod}(1963)}]{John63}
{Johnson}, H.~M. \& {MacLeod}, J.~M. 1963, \pasp, 75, 123

\bibitem[{{Kennicutt}(1984)}]{kenn84}
{Kennicutt}, R.~C. 1984, \apj, 277, 361

\bibitem[{{Kennicutt}(1998)}]{kenn98}
{Kennicutt}, R.~C. 1998, \araa, 36, 189

\bibitem[{{Li} {et~al.}(2005){Li}, {Van Dyk}, {Filippenko}, \&
  {Cuillandre}}]{li05}
{Li}, W., {Van Dyk}, S.~D., {Filippenko}, A.~V., \& {Cuillandre}, J. 2005,
  \pasp, 117, 121

\bibitem[{{Mannucci} {et~al.}(2006){Mannucci}, {Della Valle}, \&
  {Panagia}}]{mann06}
{Mannucci}, F., {Della Valle}, M., \& {Panagia}, N. 2006, \mnras, submitted
  (astro-ph/0510315)

\bibitem[{{Mannucci} {et~al.}(2005){Mannucci}, {Della Valle}, {Panagia},
  {Cappellaro}, {Cresci}, {Maiolino}, {Petrosian}, \& {Turatto}}]{mann05}
{Mannucci}, F., {Della Valle}, M., {Panagia}, N., {et~al.} 2005, \aap, 433, 807

\bibitem[{{Mannucci} {et~al.}(2003){Mannucci}, {Maiolino}, {Cresci}, {Della
  Valle}, {Vanzi}, {Ghinassi}, {Ivanov}, {Nagar}, \& {Alonso-Herrero}}]{mann03}
{Mannucci}, F., {Maiolino}, R., {Cresci}, G., {et~al.} 2003, \aap, 401, 519

\bibitem[{{Maza} \& {van den Bergh}(1976)}]{maza76}
{Maza}, J. \& {van den Bergh}, S. 1976, \apj, 204, 519

\bibitem[{{Nilson}(1973)}]{nilson}
{Nilson}, P. 1973, Uppsala general catalogue of galaxies (Acta Universitatis
  Upsaliensis.\ Nova Acta Regiae Societatis Scientiarum Upsaliensis - Uppsala
  Astronomiska Observatoriums Annaler, Uppsala: Astronomiska Observatorium,
  1973)

\bibitem[{{Patchett} \& {Wood}(1976)}]{patc76}
{Patchett}, B. \& {Wood}, R. 1976, \mnras, 175, 595

\bibitem[{{Poveda} {et~al.}(1967){Poveda}, {Ruiz}, \& {Allen}}]{pove67}
{Poveda}, A., {Ruiz}, J., \& {Allen}, C. 1967, Boletin de los Observatorios
  Tonantzintla y Tacubaya, 4, 86

\bibitem[{{Reaves}(1953)}]{reav53}
{Reaves}, G. 1953, \pasp, 65, 242

\bibitem[{{Riess} {et~al.}(1998){Riess}, {Filippenko}, {Challis},
  {Clocchiatti}, {Diercks}, {Garnavich}, {Gilliland}, {Hogan}, {Jha},
  {Kirshner}, {Leibundgut}, {Phillips}, {Reiss}, {Schmidt}, {Schommer},
  {Smith}, {Spyromilio}, {Stubbs}, {Suntzeff}, \& {Tonry}}]{riess98}
{Riess}, A.~G., {Filippenko}, A.~V., {Challis}, P., {et~al.} 1998, \aj, 116,
  1009

\bibitem[{{Riess} {et~al.}(1999){Riess}, {Kirshner}, {Schmidt}, {Jha},
  {Challis}, {Garnavich}, {Esin}, {Carpenter}, {Grashius}, {Schild}, {Berlind},
  {Huchra}, {Prosser}, {Falco}, {Benson}, {Brice{\~ n}o}, {Brown}, {Caldwell},
  {dell'Antonio}, {Filippenko}, {Goodman}, {Grogin}, {Groner}, {Hughes},
  {Green}, {Jansen}, {Kleyna}, {Luu}, {Macri}, {McLeod}, {McLeod}, {McNamara},
  {McLean}, {Milone}, {Mohr}, {Moraru}, {Peng}, {Peters}, {Prestwich},
  {Stanek}, {Szentgyorgyi}, \& {Zhao}}]{ries99}
{Riess}, A.~G., {Kirshner}, R.~P., {Schmidt}, B.~P., {et~al.} 1999, \aj, 117,
  707

\bibitem[{{Shaw}(1979)}]{shaw79}
{Shaw}, R.~L. 1979, \aap, 76, 188

\bibitem[{{Strolger} {et~al.}(2004){Strolger}, {Riess}, {Dahlen}, {Livio},
  {Panagia}, {Challis}, {Tonry}, {Filippenko}, {Chornock}, {Ferguson},
  {Koekemoer}, {Mobasher}, {Dickinson}, {Giavalisco}, {Casertano}, {Hook},
  {Blondin}, {Leibundgut}, {Nonino}, {Rosati}, {Spinrad}, {Steidel}, {Stern},
  {Garnavich}, {Matheson}, {Grogin}, {Hornschemeier}, {Kretchmer}, {Laidler},
  {Lee}, {Lucas}, {de Mello}, {Moustakas}, {Ravindranath}, {Richardson}, \&
  {Taylor}}]{stro04}
{Strolger}, L., {Riess}, A.~G., {Dahlen}, T., {et~al.} 2004, \apj, 613, 200

\bibitem[{{Valentini} {et~al.}(2003){Valentini}, {Di Carlo}, {Massi}, {Dolci},
  {Arkharov}, {Larionov}, {Pastorello}, {Di Paola}, {Benetti}, {Cappellaro},
  {Turatto}, {Pedichini}, {D'Alessio}, {Caratti o Garatti}, {Li Causi},
  {Speziali}, {Danziger}, \& {Tornamb{\' e}}}]{vale03}
{Valentini}, G., {Di Carlo}, E., {Massi}, F., {et~al.} 2003, \apj, 595, 779

\bibitem[{{van den Bergh}(1997)}]{vdb97}
{van den Bergh}, S. 1997, \aj, 113, 197

\bibitem[{{van Dyk}(1992)}]{vand92}
{van Dyk}, S.~D. 1992, \aj, 103, 1788

\bibitem[{{van Dyk} {et~al.}(1996){van Dyk}, {Hamuy}, \& {Filippenko}}]{vand96}
{van Dyk}, S.~D., {Hamuy}, M., \& {Filippenko}, A.~V. 1996, \aj, 111, 2017

\bibitem[{{van Dyk} {et~al.}(2003){van Dyk}, {Li}, \& {Filippenko}}]{vand03}
{van Dyk}, S.~D., {Li}, W., \& {Filippenko}, A.~V. 2003, \pasp, 115, 1

\bibitem[{{Zwicky}(1964)}]{zwic64}
{Zwicky}, F. 1964, \apj, 139, 514

\bibitem[{{Zwicky}(1965)}]{zwic65}
{Zwicky}, F. 1965, \pasp, 77, 456

\bibitem[{{Zwicky} {et~al.}(1963){Zwicky}, {Berger}, {Gates}, \&
  {Rudnicki}}]{zwic63}
{Zwicky}, F., {Berger}, J., {Gates}, H.~S., \& {Rudnicki}, K. 1963, \pasp, 75,
  236

\end{thebibliography}
      
\end{document}